\documentclass[12pt]{article}
\usepackage{amsmath}
\usepackage{amsfonts}
\usepackage{makeidx}
\usepackage{amssymb}
\usepackage{times}

\input{tcilatex}
\newtheorem{modass}{Assumption}
\newtheorem{notation}[theorem]{Notation}
\newtheorem{koro}[theorem]{Corollary}
\begin{document}

\title{$d$--Wave Pairing Driven by Bipolaric Modes Related to Giant
Electron--Phonon Anomalies in High--$T_{c}$ Superconductors}
\author{J.-B. Bru, A. Delgado de Pasquale and W. de Siqueira Pedra}

\maketitle

\begin{abstract}
Taking into account microscopic properties of most usual high--$T_{c}$
superconductors, like cuprates, we define a class of microscopic model
Hamiltonians for two fermions (electrons or holes) and one boson (bipolaron)
on the two--dimensional square lattice. We establish that these model
Hamiltonians can show $d$--wave paring at the bottom of their spectrum,
despite their space isotropy. This phenomenon appear when a
\textquotedblleft giant electron--phonon anomaly\textquotedblright\ is
present at the boundaries of the Brillouin zone (\textquotedblleft half
breathing\textquotedblright\ bond--stretching mode), like in doped cuprates.
Our results can be used to derive effective electron--electron interactions
mediated by bipolarons and we discuss regimes where the corresponding model
is relevant for the physics of high--temperature superconductivity and can
be mathematically rigorously studied.

\end{abstract}

\tableofcontents%

\section{Introduction}

Cuprates and many other superconducting materials with high critical
temperatures have features which are non--usual as compared to conventional
superconductors. Quoting \cite{Bonn}: \bigskip

\textit{High--temperature superconductivity in the copper oxides, first
discovered twen%
\-%
ty years ago, has led researchers on a wide-ranging quest to understand and
use this new state of matter. From the start, these materials have been
viewed as\textquotedblleft exotic\textquotedblright\ superconductors, for
which the term exotic can take on many meanings. The breadth of work that
has taken place reflects the fact that they have turned out to be exotic in
almost every way imaginable. They exhibit new states of matter (d--wave
superconductivity, charge stripes), dramatic manifestations of fluctuating
superconductivity, plus a key inspiration and testing ground for new
experimental and theoretical techniques.}\bigskip

Some of these \textquotedblleft exotic\textquotedblright\ properties, as for
instance the d--wave pairing and density waves (charge stripes) mentioned
above, turn out to be common to many high--$T_{c}$ superconductors, in
particular to those based on cuprates \cite{Bonn}. Hence, the understanding
of generic microscopic structures leading to that typical behavior can
reveal mechanisms behind high--temperature superconductivity.

In fact, the microscopic foundations of high--$T_{c}$ superconductivity are
still nowadays a subject of much debate. In the present paper we would like
to address this issue by analyzing a specific three--body problem. Indeed,
we have following aims:

\begin{itemize}
\item Taking into account microscopic properties of most usual high--$T_{c}$
superconductors (in particular cuprates), as found in recent experiments, we
define a class of \emph{microscopic} model Hamiltonians for two fermions
(electrons or holes) and one boson (bipolaron) on the two--dimensional
square lattice.

\item We mathematically rigorously analyze the spectral projection on the
bottom of the spectrum of model Hamiltonians and identify the range of
parameters that leads to $d$--wave paring.

\item We use the properties of such spectral projections in order to derive
an effective model, here called \textquotedblleft effective uncoupled
model\textquotedblright , in which the two species, bosons and fermions, do
not interact with each other.
\end{itemize}

Our main mathematical assertions are Theorems \ref{pairing mode copy(3)}, %
\ref{Carlos super main copy(2)}, \ref{Carlos super main copy(3)} and
Corollary \ref{Carlos super main copy(1)}. The paper is organized as follows:

\begin{itemize}
\item Based on experimental facts about typical high--$T_{c}$
superconductors (like cuprates), Section \ref{Section model-Hamil} gives and
discusses assumptions on model Hamiltonians.

\item Section \ref{Sectino main result} formulates the mathematical setting
and our main results. In particular, we establish that model Hamiltonians
\emph{which are invariant with respect to 90}$%
{{}^\circ}%
$\emph{--rotations} can show $d$--wave paring at the bottom of their
spectrum.

\item We derive the effective uncoupled many--body model in Section \ref%
{Sedction effective model}, using results of Section \ref{Sectino main
result}. This is reminescent of the derivation of the BCS model where the
attraction between electrons is mediated by bosonic degrees of freedom (e.g.
phonons). See also the Fr\"{o}hlich model. The way we do it is however
different from the usual derivations and is mathematically rigorous in the
2-fermions--1-boson sector. Moreover, this new model can explain $d$--wave
superconductivity, in contrast to the BCS model\footnote{%
Note that many authors contributed to the mathematics of the (usual)
BCS-model. For instance, to mention some of them: Haag, Thirring, Lieb,
Bogoliubov (Jr), Zagrebnov, or more recently Seiringer and Hainzl, etc.}.

\item Section \ref{interesting sect proofs} gathers technical proofs on
which Sections \ref{Sectino main result}--\ref{Sedction effective model} are
based.

\item Section \ref{appendix} is an appendix on direct integral
decompositions and the Birman--Schwinger principle, which are important
technical tools to prove our assertions.
\end{itemize}

\begin{notation}
\label{remark constant copy(1)}\mbox{
}\newline
To simplify notation, we denote positive and finite constants by $D\in
(0,\infty )$. These constants do not need to be the same from one statement
to another. We denote the Banach space of bounded operators acting on a
Hilbert space $\mathcal{H}$ by $\mathcal{B}(\mathcal{H})$ with operator norm
$\Vert \cdot \Vert _{\mathrm{op}}$ and identity $\mathbf{1}_{\mathcal{H}}$.
\end{notation}

\section{Prototypical Properties of High--$T_{c}$ Superconductors\label%
{Section model-Hamil}}

In the next four subsections we briefly discuss some experimental facts
giving, in our opinion, important hints about the nature of the microscopic
interaction involving electrons in superconducting cuprates. Based on this
discussion, we propose a list of assumptions on our model Hamiltonians.

\subsection{Electron Repulsion and Hoppings}

It is well-known that undopped cuprates are insulators. Moreover,
experiments showed that the insulating phase of cuprates is indeed a
so--called \textquotedblleft Mott insulating phase\textquotedblright . See
for instance \cite{Bonn} for a review. This phase is characterized by a
periodic distribution of electrons with exactly one particle per lattice
site. Such a space distribution of electrons is a consequence of a strong
repulsion of two charge particles sitting at the same lattice site. Dopping
cuprates with holes or electrons leads to a mean density $\rho $ different
from one electron per site and the above configuration is not anymore
energetically favorable. It turns out that, in this case, at sufficiently
small temperatures, the superconducting phase is the one minimizing the
free--energy density. In particular, the system undergoes a phase transition
and becomes a superconductor. This phenomenon was rigorously proven in \cite%
{BruPedra1} for the strong coupling reduced BCS Hamiltonian perturbed by a
repulsive Hubbard interaction. Further properties of the phase diagram of
real cuprates are also captured if we consider the two--band version of this
Hamiltonian. In fact, for real cuprates the phase diagram is not symmetric
with respect to the axis $\rho =1$ (no doping, one electron per site). The
critical temperature tends to be higher for hole doping than for electron
doping. This property is shown to be true for the two--band model studied in
\cite{BruPedraAniko}.

The results of \cite{BruPedra1,BruPedraAniko} confirm, from a mathematical
point of view, that the shape of the typical phase diagram of cuprates as
well as the corresponding type of phase transition can be drawn back to the
competition between a strongly repulsive short--range force between
electrons and a weak but long--range BCS--type interaction. We thus assume
the following:

\begin{modass}[Hubbard repulsion]
\label{MArep}\mbox{ }\newline
The repulsive force between two near--lying electrons is represented by the
usual Hubbard repulsive interaction (which does not vanish only for
particles at the same lattice site).
\end{modass}

The absence of hopping terms in the Hamiltonian studied in \cite%
{BruPedra1,BruPedraAniko} corresponds to the so--called \textquotedblleft
strong coupling approximation\textquotedblright\ for the BCS model. Here, we
aim to introduce general hopping terms in our models. Note however that the
\textquotedblleft strong coupling regime\textquotedblright\ is, from one
side, technically convenient, but, first of all, also the most relevant case
in which concerns high--$T_{c}$ superconductivity: Experiments suggest \cite%
{Saxena} that the inter--particle interaction energy is five to eight times
bigger than the hopping strength:

\begin{modass}[Strong coupling regime]
\label{MArep copy(1)}\mbox{ }\newline
The interactions between particles are strong with respect to the hopping
amplitudes.
\end{modass}

Charge transport in cuprates take place within separated (almost)
independent layers. In fact, we focus on high--$T_{c}$ materials for which
superconducting carriers, mainly holes in the case of cuprates, move within
two--dimensional $\mathrm{CuO}_{2}$ layers made of $\mathrm{Cu}^{++}$ and $%
\mathrm{O}^{--}$, see, e.g., \cite[Fig. 5.3. p. 127]{Saxena}. The following
assumption is thus reasonable:

\begin{modass}[Two--dimensionality]
\label{MA2D}\mbox{ }\newline
The charge transport occurs within independent two--dimensional layers.
\end{modass}

We also know from \cite{BruPedra1,BruPedraAniko} that the reduced BCS
interaction, also in presence of the Hubbard repulsive term, always lead to $%
s$--wave pairing of electrons in the superconducting phase. Hence, this
component should be replaced by another effective long--range attractive
force. Effective microscopic forces between electrons, which could play a
role in the phenomenon of $d$--wave pairing, are deduced in Section \ref%
{Sedction effective model} from results of Section \ref{Sectino main result}%
. One important physical fact that gives hint on the nature of the
microscopic forces leading to high--temperature superconductivity are the
anomalous dispersion relations of phonons in high--$T_{c}$ superconductors
discussed in the following subsection.

\subsection{Giant Electron--Phonon Anomalies in Dopped Cuprates}

Anomalous dispersion relations of phonons, i.e., dispersion relations of a
form not expected from lattice--dynamical models, are usually due to the
coupling of phonon to electrons. Such a phenomenon is observed even in
conventional metals. In doped cuprates and other high--$T_{c}$
superconductors such an anomaly is very strong (\textquotedblleft giant
electron--phonon anomaly\textquotedblright ) and very localized in specific
regions of the Brillouin zone, suggesting a strong interaction between
elastic and charged modes in some small range of quasi--momenta. For a
recent review on giant electron--phonon anomalies in doped cuprates see for
instance \cite{Rez}. Experiments with cuprates show that these anomalies get
stronger at the boundaries of the Brillouin zone (\textquotedblleft half
breathing\textquotedblright\ bond--stretching mode) as the doping is
increased \cite{28}. Since, until a certain point, the increasing of doping
also increases the superconducting critical temperature, it is natural to
expect that a strong coupling between charged and \textquotedblleft half
breathing\textquotedblright\ bond-stretching modes is part of the mechanism
leading to high temperature superconductivity \cite{29}. In a
two--dimensional model for superconductors \textquotedblleft half
breathing\textquotedblright\ bond--stretching modes correspond to $(\pm \pi
,0)$ and $(0,\pm \pi )$ quasi--momentum transfers.

The precise type of coupling between charged and elastic modes responsible
for the giant electron--phonon anomalies is a subject of debate. One
possible mechanism is the existence incipient instability due to the
formation of polarons and bipolarons \cite{61,62,63,64,65}. For other
mechanisms see the review \cite{Rez}. In the next subsection we discuss the
bipolaronic scenario in more details. The following physical assumption,
that is, strong bipolaron instabilities at quasi--momenta $(\pm \pi ,0)$ and
$(0,\pm \pi )$, is made with respect to the two--dimensional microscopic
models we consider:

\begin{modass}[Strong bipolaron instabilities]
\label{MAbip}\mbox{ }\newline
The strong interaction between elastic and charged modes at half breathing
bond--stretching modes is related to the formation of bipolarons.
\end{modass}

We also assume the following condition:

\begin{modass}[Zero--spin bipolarons]
\label{MAzspin}\mbox{ }\newline
The total spin of bipolarons is zero.
\end{modass}

\noindent Considering spin--one bipolarons would also be feasible, but we
refrain from doing it for simplicity.

\subsection{Bipolaron--Electron Exchange Interaction}

There are experimental evidences of polaron and bipolaron formation in high--%
$T_{c}$ superconductors, even in insulating and metallic phases. See, for
instance, \cite{Ranninger} for a brief review on these experimental issues.
Some efforts have been made to theoretically explain high--$T_{c}$
superconductivity by assuming that bipolarons, and not Cooper pairs, are the
main charge carriers in the superconducting phase \cite{1-0,1-1}. Recall
that it is experimentally known that, for cuprates and other high--$T_{c}$
superconductors, the charge carriers in the superconducting phase have two
times the charge of the electron, as in the case of conventional
superconductors. Nevertheless, there is an important objection to this
picture: polarons and bipolarons (more generally, $n$--polarons, $n\in
\mathbb{N}$) are charge carriers that are self--trapped inside a strong and
local lattice deformation that surounds them (their are electrons
\textquotedblleft dressed with phonons\textquotedblright ). Such strong
lattice deformations attached to bipolarons can hardly move and this is not
in accordance with the known mobility of superconducting charge carriers.
Hence we assume:

\begin{modass}[Small bipolaron mobility]
\label{MAbipmob}\mbox{ }\newline
The hopping strength of bipolarons (bosonic particles) is very small or even
negligible.
\end{modass}

One way out of this mobility problem is to assume that bipolarons can decay
into two--electrons and, reciprocally, two moving electrons can bind
together to form a new bipolaron \cite{Ranninger}:

\begin{modass}[Bipolaron--electron exchange]
\label{MAext}\mbox{ }\newline
Bipolarons can decay into two electrons and moving electrons can bind to
form bipolarons. This exchange process is strong for quasi--momenta near the
half breathing bond--stretching modes, i.e., in two--dimensions, $(\pm \pi
,0)$, $(0,\pm \pi )$, and weak away from these singular points.
\end{modass}

This exchange process allows a good mobility of charge carriers because the
electronic state has a non--negligible hopping strength. Moreover, such a
boson--fermion exchange process effectively creates an attractive force
between electrons, like in the Fr\"{o}hlich model for conventional
superconductivity. So, we expect a binding mechanism for electron pairs
similar, in a sense, to the Cooper pairing, but with the mediating boson
being a bipolaron instead of (directly) a phonon.

As the exchange process described above is concentrated (in momentum space)
around a few isolated points (half breathing bond-stretching modes) of the
Brillouin zone, it is conceivable that the following holds true:

\begin{modass}[Long--range effective forces]
\label{MAlr}\mbox{ }\newline
The forces between electrons mediated by bipolarons are long--ranged (in
space).
\end{modass}

This assumption is not in contradiction with the experimentally known fact
that the pairs responsible for charge transport in high--$T_{c}$
superconductors have (in contrast to conventional superconductors) a very
small extension. Indeed, as shown in \cite{BruPedra1}, the small space
extension of superconducting pairs is rather due to the strong coupling
regime. Note moreover that Assumption \ref{MAlr} is not used in Section \ref%
{Sectino main result}. It is only relevant for the effective many--body
model we propose in Section \ref{Sedction effective model}.

\subsection{Space Isotropy}

There are theoretical studies showing that an anisotropic phonon--electron
(or, more generally, boson--fermion) interaction can explain $d$--wave
pairing of electrons \cite{58}. On the other hand, there is absolutely no
evidence of such an anisotropy in cuprates, see \cite{Rez} for instance. So,
we aim to derive $d$--pairing (among other phenomena typical to high--$T_{c}$
superconductors) from strictly isotropic models and we assume the following:

\begin{modass}[Isotropy of interactions]
\label{MAinv}\mbox{ }\newline
The interactions are invariant unter lattice translations, reflections and 90%
$%
{{}^\circ}%
$--rota\textit{%
\-%
}tions.
\end{modass}

This condition concludes the list of assumptions on which we base our
mathematically rigorous study.

\section{Mathematical Setting and Main Results\label{Sectino main result}}

In this section, we mathematically implement (the physical) Assumptions \ref%
{MArep}--\ref{MAext} and \ref{MAinv}.

\subsection{Bipolaron--Electron Model for High--$T_{c}$ Superconductors}

By taking into account all model assumptions formulated above, we propose
bellow a Hamiltonian for bosons and fermions in the $\mathbb{Z}^{2}$%
--lattice. In particular, the host material supporting particles is assumed
to be a (perfect) two--dimensional cubic crystal (cf. Assumption \ref{MA2D}).

For any $n\in \mathbb{N}$, let $\mathcal{S}_{n,\pm }$ be the orthogonal
projections onto the subspace of, respectively, antisymmetric ($-$) and
symmetric ($+$) $n$--particle wave functions in $\mathfrak{h}_{\pm
}^{\otimes n}$, the $n$--fold tensor product of either $\mathfrak{h}%
_{-}:=\ell ^{2}(\mathbb{Z}^{2};\mathbb{C}^{2})$ or $\mathfrak{h}_{+}:=\ell
^{2}(\mathbb{Z}^{2};\mathbb{C})$. Let $\mathfrak{h}_{n,\pm }:=\mathcal{S}%
_{n,\pm }\mathfrak{h}_{\pm }^{\otimes n}$ for all $n\in \mathbb{N}$, $%
\mathfrak{h}_{0,\pm }:=\mathbb{C}$, and define%
\begin{equation*}
\mathcal{F}_{\pm }:=\bigoplus_{n=0}^{\infty }\mathfrak{h}_{n,\pm }
\end{equation*}%
to be respectively the fermion (($-$), spin $1/2$) and boson (($+$),
spinless, cf. Assumptions \ref{MAbip} and \ref{MAzspin}) Fock spaces. The
Hilbert space of the compound system is thus
\begin{equation*}
\mathcal{F}_{-,+}:=\mathcal{F}_{-}\otimes \mathcal{F}_{+}\simeq
\bigoplus_{n,m=0}^{\infty }\mathfrak{h}_{m,-}\otimes \mathfrak{h}_{n,+}\ .
\end{equation*}%
Here, $\simeq $ denotes the existence of a \emph{canonical} isomorphism of
Hilbert spaces. A dense subset of $\mathcal{F}_{-,+}$ is given by the
subspace
\begin{equation}
\mathcal{D}:=\mathrm{span}\left\{ \bigcup\limits_{m,n\in \mathbb{N}_{0}}%
\mathfrak{h}_{m,-}\otimes \mathfrak{h}_{n,+}\right\} \ .
\label{dense subspace}
\end{equation}

The creation and annihilation operators are denoted by%
\begin{equation*}
a_{x,\mathrm{s}}^{\ast }\equiv a_{x,\mathrm{s}}^{\ast }\otimes \mathbf{1}_{%
\mathcal{F}_{+}}\ ,\ \ a_{x,\mathrm{s}}\equiv a_{x,\mathrm{s}}\otimes
\mathbf{1}_{\mathcal{F}_{+}}\ ,\ \ x\in \mathbb{Z}^{2},\ \mathrm{s}\in
\{\uparrow ,\downarrow \}\ ,
\end{equation*}%
for fermions and
\begin{equation*}
b_{x}^{\ast }\equiv \mathbf{1}_{\mathcal{F}_{-}}\otimes b_{x}^{\ast }\
,\quad b_{x}\equiv \mathbf{1}_{\mathcal{F}_{-}}\otimes b_{x}\ ,\quad x\in
\mathbb{Z}^{2},
\end{equation*}%
in the boson case.

The fermionic part of the (infinite volume) Hamiltonian is defined on the
dense subspace $\mathcal{D}\subset \mathcal{F}_{-,+}$ by the symmetric
operator%
\begin{equation}
H_{f}:=\epsilon \left( -\frac{1}{2}\underset{\mathrm{s}\in \{\uparrow
,\downarrow \},x,y\in \mathbb{Z}^{2},|x-y|=1}{\sum }a_{x,\mathrm{s}}^{\ast
}a_{y,\mathrm{s}}+2\underset{\mathrm{s}\in \{\uparrow ,\downarrow \},x\in
\mathbb{Z}^{2}}{\sum }a_{x,\mathrm{s}}^{\ast }a_{x,\mathrm{s}}\right) +U%
\underset{x\in \mathbb{Z}^{2}}{\sum }n_{x,\uparrow }n_{x,\downarrow }
\label{interesting hamiltonian free electrons}
\end{equation}%
with $\epsilon ,U\geq 0$. The first term of this operator represents, as
usual, next--neighbor hoppings of fermions on the $\mathbb{Z}^{2}$--lattice.
More generally, we could take a term of the form
\begin{equation*}
\epsilon \underset{\mathrm{s}\in \{\uparrow ,\downarrow \},x,y\in \mathbb{Z}%
^{2}}{\sum }h_{f}(|x-y|)a_{x,\mathrm{s}}^{\ast }a_{y,\mathrm{s}}\ ,
\end{equation*}%
for some real--valued function $h_{f}$ satisfying
\begin{equation*}
\sum\limits_{x\in \mathbb{Z}^{2}}|h_{f}(x)|<\infty \ .
\end{equation*}%
We refrain from considering this general case for simplicity, only, but our
study can be easily generalized to this situation. The last term of (\ref%
{interesting hamiltonian free electrons}) stands for the (screened) Coulomb
repulsion as in the celebrated Hubbard model. So, the parameter $U$ is a
positive number, i.e., $U\geq 0$. See Assumption \ref{MArep}. The parameter $%
\epsilon \geq 0$ represents the relative strengh of the hopping amplitude
with respect to the interparticle interaction. In high--$T_{c}$
superconductors, $\epsilon $ is expected to be relatively small. Cf.
Assumption \ref{MArep copy(1)}.

The bosonic part of the Hamiltonian is meanwhile defined on $\mathcal{D}$ by%
\begin{equation}
H_{b}:=\epsilon \left( -\frac{h_{b}}{2}\underset{x,y\in \mathbb{Z}%
^{2},|x-y|=1}{\sum }b_{x}^{\ast }b_{y}+2h_{b}\underset{x\in \mathbb{Z}^{2}}{%
\sum }b_{x}^{\ast }b_{x}\right) \,,  \label{boson hamiltonian}
\end{equation}%
where $h_{b}\geq 0$ is very small or even zero (cf. Assumption \ref{MAbipmob}%
). This symmetric operator does not include any density--density
interaction. Indeed, we only consider below the one--boson subspace and such
interactions are thus irrelevant in the sequel. [If some density--density
interaction is added here for the bosons, then the effective model (\ref%
{effective boson model}) has to include it.]

We define the full Hamiltonian (fermion--boson compound system) by the
symmetric operator%
\begin{equation}
\mathbf{H}:=H_{f}+H_{b}+W\in \mathcal{L}(\mathcal{D},\mathcal{F}_{-,+})\ ,
\label{interaction W0}
\end{equation}%
where $\mathcal{L}(\mathcal{D},\mathcal{F}_{-,+})$ stands for the space of
linear operators from $\mathcal{D}$ to $\mathcal{F}_{-,+}$ and
\begin{equation}
W:=\underset{x,y\in \mathbb{Z}^{2}}{\sum }v\left( x-y\right) \left(
b_{x}^{\ast }c_{y}+c_{y}^{\ast }b_{x}\right)  \label{interaction W}
\end{equation}%
encodes (spin--conserving) exchange interactions between electron pairs and
bipolarons (cf. Assumption \ref{MAext}) with $\mathbb{Z}^{2}$--summable
coupling functions $v$. The fermionic operator $c_{x}$ is defined, for all $%
x\in \mathbb{Z}^{2}$ and some (large) parameter $\kappa >0$, by%
\begin{equation}
c_{x}:=\underset{z\in \mathbb{Z}^{2},\left\vert z\right\vert \leq 1}{\sum }%
\mathrm{e}^{-\kappa \left\vert z\right\vert }a_{x+z,\uparrow
}a_{x,\downarrow }\ .  \label{cooper pair2}
\end{equation}

Observe that the fermion--boson coupling is completely different from the
one used in the celebrated Fr\"{o}hlich model. For instance, in contrast to
the Fr\"{o}hlich model, it does not conserve the fermion number. Note also
that we do not use the operator
\begin{equation}
\tilde{c}_{x}:=\underset{z\in \mathbb{Z}^{2}}{\sum }\mathrm{e}^{-\kappa
\left\vert z\right\vert }a_{x+z,\uparrow }a_{x,\downarrow }=c_{x}+\mathcal{O}%
(\mathrm{e}^{-\sqrt{2}\kappa })  \label{cooper pair1}
\end{equation}%
instead of $c_{x}$ in the sequel in order to simplify technical arguments,
only. The action of $\tilde{c}_{x}$ can be viewed as the annihilation of an
electron pair localized in a region of radius $\mathcal{O}(\kappa ^{-1})$.
It could be interesting to replace the Hubbard repulsion by general
density--density interaction resulting from the second quantization of
two--body interactions, like for instance
\begin{equation}
U\sum\limits_{\mathrm{s}\in \{\uparrow ,\downarrow \},x,y\in \mathbb{Z}%
^{2}}u\left( \left\vert x-y\right\vert \right) a_{y}^{\ast }a_{y}a_{x}^{\ast
}a_{x}  \label{density-density extension}
\end{equation}%
on $\mathcal{D}$, where $u\left( r\right) :\mathbb{R}_{0}^{+}\rightarrow
\mathbb{R}^{+}$ is some real--valued function. See discussion at the end of
Section \ref{section main}.

We consider fermion--boson interactions (\ref{interaction W}) with
real--valued coupling functions $v$ which are $\mathbb{Z}^{2}$--summable,
symmetric and 90$%
{{}^\circ}%
$--rotation invariant (cf. Assumption \ref{MAinv}): $v\in \ell ^{1}\left(
\mathbb{Z}^{2},\mathbb{R}\right) $ and, for all $x\equiv (x_{1},x_{2})\in
\mathbb{Z}^{2}$,
\begin{equation*}
v(x)=v(-x)\ ,\quad v(x)\equiv v(x_{1},x_{2})=v(-x_{2},x_{1})\ .
\end{equation*}%
Note that the Fourier transform $\hat{v}$ of such a $v$ exists as a
real--valued continuous function which is symmetric and 90$%
{{}^\circ}%
$--rotation invariant, i.e., for all $k\equiv (k_{1},k_{2})\in \lbrack -\pi
,\pi )^{2}$,
\begin{equation}
\hat{v}(k)=\hat{v}(-k)\ ,\quad \hat{v}(k)\equiv \hat{v}(k_{1},k_{2})=\hat{v}%
(-k_{2},k_{1})\ .  \label{because of that}
\end{equation}%
It is convenient, for reasons which become clear later on, to take $v$ of
the form
\begin{equation}
v=v_{+}-v_{-}\ ,  \label{decompositino}
\end{equation}%
where $v_{\pm }\in \ell ^{1}\left( \mathbb{Z}^{2},\mathbb{R}\right) $ are
functions of positive type, i.e., their Fourier transforms $\hat{v}_{\pm }$
are non--negative.

\subsection{$d$--Wave Pairing in the 2-Fermions--1-Boson Sector\label%
{section main}}

We aim to study the unitary group generated by $\mathbf{H}$ on the smallest
invariant space of $\mathbf{H}$ containing the subspace related to one pair
of electrons with total spin equal to zero. This invariant space is%
\begin{equation}
\mathcal{H}_{\uparrow ,\downarrow }^{(2,1)}:=(\mathfrak{h}%
_{2,-}^{(0)}\otimes \mathfrak{h}_{0,+})\oplus (\mathfrak{h}_{0,-}\otimes
\mathfrak{h}_{1,+})\simeq \mathfrak{h}_{2,-}^{(0)}\oplus \mathfrak{h}_{1,+}
\label{hilbert space elementaire1}
\end{equation}%
with $\mathfrak{h}_{2,-}^{(0)}$ being the subspace of one zero--spin fermion
pair. $\mathfrak{h}_{2,-}^{(0)}$ is canonically isomorphic to the spaces%
\begin{equation}
\ell ^{2}\left( \mathbb{Z}^{2};\mathbb{C}\right) \otimes \ell ^{2}\left(
\mathbb{Z}^{2};\mathbb{C}\right) \simeq \ell ^{2}(\mathbb{Z}^{2}\times
\mathbb{Z}^{2};\mathbb{C})\ .  \label{cooper pair space}
\end{equation}%
The first Hilbert space $\ell ^{2}\left( \mathbb{Z}^{2};\mathbb{C}\right) $
in the tensor product encodes the wave functions of a fermion with spin up ($%
\uparrow $), whereas the second one refers to a fermion with spin down ($%
\downarrow $). The isomorphism between $\mathfrak{h}_{2,-}^{(0)}$ and $\ell
^{2}(\mathbb{Z}^{2}\times \mathbb{Z}^{2};\mathbb{C})$ is choosen in such a
way that, for any $x,y\in \mathbb{Z}^{2}$, $a_{x,\uparrow }a_{y,\downarrow }$%
, seen as an operator from $\ell ^{2}(\mathbb{Z}^{2}\times \mathbb{Z}^{2};%
\mathbb{C})$ to $\mathbb{C}$, satisfies
\begin{equation}
a_{x,\uparrow }a_{y,\downarrow }\left( \mathfrak{c}\right) =\mathfrak{c}%
\left( x,y\right) \ ,\qquad \mathfrak{c}\in \ell ^{2}(\mathbb{Z}^{2}\times
\mathbb{Z}^{2};\mathbb{C})\ .  \label{idiot}
\end{equation}%
Since $\mathfrak{h}_{1,+}=\ell ^{2}(\mathbb{Z}^{2};\mathbb{C})$, it follows
that%
\begin{equation}
\mathcal{H}_{\uparrow ,\downarrow }^{(2,1)}\simeq \ell ^{2}(\mathbb{Z}%
^{2}\times \mathbb{Z}^{2};\mathbb{C})\times \ell ^{2}(\mathbb{Z}^{2};\mathbb{%
C})\ .  \label{hilbert space elementaire2}
\end{equation}%
In particular, we denote elements $\psi $ of $\mathcal{H}_{\uparrow
,\downarrow }^{(2,1)}$ by $\psi =(\mathfrak{c},\mathfrak{b})$, where $%
\mathfrak{c}$ and $\mathfrak{b}$ are respectively the wave function of one
fermion pair and one boson. Observe that
\begin{equation}
H^{(2,1)}:=\overline{\mathbf{H}|_{\mathcal{H}_{\uparrow ,\downarrow
}^{(2,1)}}}  \label{important hamiltonian boson fermi}
\end{equation}%
is a bounded self--adjoint operator on the subspace $\mathcal{H}_{\uparrow
,\downarrow }^{(2,1)}$. Recall that $v\in \ell ^{1}\left( \mathbb{Z}^{2},%
\mathbb{R}\right) $. We study below the unitary group generated by the
Hamiltonian $H^{(2,1)}$ in order to show the formation of a\ bound fermion
pair of minimum energy via a mediating bipolaron (spinless boson, in the
present case), as discussed in the introduction.

To this end, we define the ground state energy of the Hamiltonian $H^{(2,1)}$
by
\begin{equation*}
E_{0}:=\inf \sigma (H^{(2,1)})\ ,
\end{equation*}%
where $\sigma (A)$ is, by definition, the spectrum of any self--adjoint
operator $A$. From Lemma \ref{Carlos super main copy(5)}, observe that $%
E_{0}\leq 0$. In fact, we can give an explicit criterium for the strict
negativity of $E_{0}$, which is interpreted as a \emph{bound fermion pair }%
formation. See Theorem \ref{Carlos super main copy(2)} and discussion
thereafter.

Indeed, for any $\epsilon \geq 0$, $\kappa >0$, $\lambda <0$ and $k\in
\lbrack -\pi ,\pi )^{2}$, let%
\begin{eqnarray}
R_{\mathfrak{s},\mathfrak{s}}^{(0)} &:=&\frac{1}{(2\pi )^{2}}\int_{[-\pi
,\pi )^{2}}\frac{1}{\epsilon \left( 4-\cos (k_{\uparrow \downarrow }-k)-\cos
(k_{\uparrow \downarrow })\right) -\lambda }\ \mathrm{d}^{2}k_{\uparrow
\downarrow }\ ,  \label{Rss} \\
R_{\mathfrak{d},\mathfrak{d}}^{(0)} &:=&\frac{1}{(2\pi )^{2}}\int_{[-\pi
,\pi )^{2}}\frac{(1+2\mathrm{e}^{-\kappa }\cos (k_{\uparrow \downarrow
}-k))^{2}}{\epsilon \left( 4-\cos (k_{\uparrow \downarrow }-k)-\cos
(k_{\uparrow \downarrow }\right) )-\lambda }\ \mathrm{d}^{2}k_{\uparrow
\downarrow }\ , \\
R_{\mathfrak{s},\mathfrak{d}}^{(0)} &:=&\frac{1}{(2\pi )^{2}}\int_{[-\pi
,\pi )^{2}}\frac{1+2\mathrm{e}^{-\kappa }\cos (k_{\uparrow \downarrow }-k)}{\epsilon \left( 4-\cos (k_{\uparrow \downarrow }-k)-\cos (k_{\uparrow
\downarrow }\right) )-\lambda }\ \mathrm{d}^{2}k_{\uparrow \downarrow }\ ,
\label{Rss2}
\end{eqnarray}%
where
\begin{equation}
\cos (q):=\cos (q_{x})+\cos (q_{y})\ ,\qquad q\equiv (q_{x},q_{y})\in
\lbrack -\pi ,\pi )^{2}\ .  \label{cosinus}
\end{equation}%
These strictly positive constants can easily be determined to a very high
precision by numerical computations. Then, define the (possibly infinite)
numbers%
\begin{eqnarray*}
\mathrm{I}(k,U) &:=&\ \underset{\lambda \rightarrow 0^{-}}{\lim \sup }\left\vert \hat{v}\left( k\right) \right\vert ^{2}\left\{ \frac{R_{\mathfrak{d},\mathfrak{d}}^{(0)}}{1+UR_{\mathfrak{s},\mathfrak{s}}^{(0)}}+U\frac{R_{\mathfrak{d},\mathfrak{d}}^{(0)}R_{\mathfrak{s},\mathfrak{s}}^{(0)}-\left(
R_{\mathfrak{s},\mathfrak{d}}^{(0)}\right) ^{2}}{1+UR_{\mathfrak{s},\mathfrak{s}}^{(0)}}\right\} \in \left[ 0,\infty \right]  \\
\mathrm{I}(k,\infty ) &:=&\ \underset{\lambda \rightarrow 0^{-}}{\lim \sup }\left\vert \hat{v}\left( k\right) \right\vert ^{2}\left\{ \frac{R_{\mathfrak{d},\mathfrak{d}}^{(0)}R_{\mathfrak{s},\mathfrak{s}}^{(0)}-\left( R_{\mathfrak{s},\mathfrak{d}}^{(0)}\right) ^{2}}{R_{\mathfrak{s},\mathfrak{s}}^{(0)}}\right\} \in \left[ 0,\infty \right]
\end{eqnarray*}%
for any $U\geq 0$ and $k\in \lbrack -\pi ,\pi )^{2}$, see (\ref{postivity}).
A sufficient condition to obtain a bound pair (i.e., $E_{0}<0$) is as
follows:

\begin{theorem}[Strict negativity of $E_{0}$]
\label{pairing mode copy(3)}\mbox{ }\newline
\emph{(i)} $E_{0}<0$ if and only if%
\begin{equation*}
\sup_{k\in \lbrack -\pi ,\pi )^{2}}\left\{ \mathrm{I}(k,U)-\epsilon
h_{b}(2-\cos (k))\right\} >0\ .
\end{equation*}%
The latter always holds true whenever $\hat{v}\left( 0\right) \neq 0$.
\newline
\emph{(ii)} At fixed $\epsilon \geq 0$,%
\begin{equation*}
\underset{U\rightarrow \infty }{\lim \inf }E_{0}<0
\end{equation*}%
if and only if%
\begin{equation*}
\sup_{k\in \lbrack -\pi ,\pi )^{2}}\left\{ \mathrm{I}(k,\infty )-\epsilon
h_{b}(2-\cos (k))\right\} >0\ .
\end{equation*}
\end{theorem}

\noindent \textit{Proof.} The assertions are direct consequences of
Proposition \ref{effective BCS copy(1)} and Lemmata \ref{effective BCS
copy(4)}, \ref{effective BCS copy(2)bis} and \ref{effective BCS copy(2)+1}.
\hfill $\Box $

\noindent Note that Theorem \ref{pairing mode copy(3)} yields $E_{0}<0$ for
sufficiently small $\epsilon \geq 0$, unless $v=0$.

We are interested in the time evolution driven by this Hamiltonian for
three--body wave functions with minimum energy. We thus consider initial
wave functions $(\mathfrak{c}_{0},\mathfrak{b}_{0})$ in the subspace%
\begin{equation}
\mathfrak{H}_{\varepsilon }:=\mathrm{Ran}\left( \mathbf{1}%
_{[E_{0},E_{0}\left( 1-\varepsilon \right) ]}(H^{(2,1)})\right) \subset
\mathcal{H}_{\uparrow ,\downarrow }^{(2,1)}\
\label{interesting spectral space}
\end{equation}%
for small $\varepsilon >0$. Here, for any $\alpha _{1},\alpha _{2}\in
\mathbb{R}$, $\alpha _{1}<\alpha _{2}$, and self--adjoint operator $A$, $%
\mathbf{1}_{[\alpha _{1},\alpha _{2}]}(A)$ denotes the spectral projector of
$A$ associated to its spectrum in the interval $[\alpha _{1},\alpha _{2}]$,
while $\mathrm{Ran}(A)$ stands for the range of $A$. Note that $E_{0}$ is
generally not an eigenvalue of $H^{(2,1)}$, see Section \ref{interesting
sect proofs}.

Then, for any positive real number $0<\varepsilon \ll 1$, we study the
properties of the time--dependent wave function $(\mathfrak{c}_{t},\mathfrak{%
b}_{t})$, solution of the Schr\"{o}dinger equation%
\begin{equation}
\forall t\in \mathbb{R}:\ \ i\frac{\mathrm{d}}{\mathrm{d}t}(\mathfrak{c}_{t},%
\mathfrak{b}_{t})=H^{(2,1)}(\mathfrak{c}_{t},\mathfrak{b}_{t})\ ,\ \ (%
\mathfrak{c}_{t},\mathfrak{b}_{t})\in \mathfrak{H}_{\varepsilon }\ .
\label{le chat}
\end{equation}%
With this aim, define the (non--empty) set%
\begin{equation*}
\Xi _{\varepsilon }:=\Big \{(\mathfrak{c}_{t},\mathfrak{b}_{t})\in C^{1}(%
\mathbb{R},\mathfrak{H}_{\varepsilon })\,\text{norm--one solution of
Equation (\ref{le chat})}\Big \}
\end{equation*}%
for $0<\varepsilon \ll 1$. Units are chosen so that $\hbar =1$.

We first show that the strict negativity of $E_{0}$ corresponds to the
existence a bound fermion pair:

\begin{theorem}[Existence of bound fermion pairs]
\label{Carlos super main copy(2)}\mbox{ }\newline
Assume that $E_{0}<0$. For any $\eta ,\varepsilon \in (0,1)$ and $(\mathfrak{%
c}_{t},\mathfrak{b}_{t})\in \Xi _{\varepsilon }$, there is a constant $%
R<\infty $ such that, for all $t\in \mathbb{R}$,%
\begin{equation*}
\sum_{x_{\uparrow },x_{\downarrow }\in \mathbb{Z}^{2}:|x_{\uparrow
}-x_{\downarrow }|\leq R}|\mathfrak{c}_{t}(x_{\uparrow },x_{\downarrow
})|^{2}\geq \Vert \mathfrak{c}_{0}\Vert _{2}^{2}(1-\eta )>0\ .
\end{equation*}%
Moreover, if the hopping amplitude $\epsilon >0$ is sufficiently small then
one can choose $R=1$.
\end{theorem}

\noindent \textit{Proof.} It is a direct consequence of Proposition \ref%
{effective BCS copy(7)}.\hfill $\Box $

Since $(\mathfrak{c}_{t},\mathfrak{b}_{t})\in \Xi _{\varepsilon }$ has norm
one for all $t\in \mathbb{R}$, we infer from Theorem \ref{Carlos super main
copy(2)} that, uniformly in time $t$, the probability of finding an electron
pair in a region of diameter $1$ is always strictly positive for strictly
negative $E_{0}<0$ and sufficiently small hopping amplitude $0\leq \epsilon
\ll 1$. In other words, the fermion part $\mathfrak{c}_{t}$ never vanishes
in this regime while the two fermions behave as a composite particle, i.e.,
a bound fermion pair. We also have a non--vanishing boson part $\mathfrak{b}%
_{t}$ for all times. The latter can be seen as a depletion of either the
pair density or the boson (bipolaron) density. This depletion results from
the interaction $W$ (\ref{interaction W}) which implies an effective
attraction between fermions. This can heuristically be understood by
diagrammatic methods like in \cite{BZ1}. It is also reminiscent of the Bose
condensate depletion found in the rigorous study of the Bogoliubov model and
its variants. See for instance \cite{BruZagrebnov8,bru3}. The boson--fermion
occupation ratio can be explicitly computed in the limits $\epsilon
\rightarrow 0$ and $U\rightarrow \infty $:
\begin{equation}
\frac{\Vert \mathfrak{c}_{t}\Vert _{2}^{2}}{\Vert \mathfrak{c}_{t}\Vert
_{2}^{2}+\Vert \mathfrak{b}_{t}\Vert _{2}^{2}}\rightarrow \frac{1}{2}
\label{depletion}
\end{equation}%
see Lemma \ref{effective BCS copy(6)}. Similar results in the regime $%
\epsilon \rightarrow 0$ and $U\rightarrow 0$ can also be deduced from our
study.

For sufficiently small hopping amplitudes, the last theorem says that the
bound pair is (s) either localized on a single lattice site or (d) the
fermions forming the\ bound pair have distance exactly equal to $1$ to each
other. (s) mainly appears at small coupling $U\geq 0$ and corresponds to a $%
s $--wave pair. By contrast, (d) occurs at large $U\geq 0$ and is related to
the formation of a $d$--wave pair whenever Assumption \ref{MAext} is
satisfied. If this assumption does not holds, we still have, at large $U\geq
0$, a distance exactly equal to $1$ between the fermions in the bound pair,
but the pairing symmetry is rather of \emph{generalized} $s$--wave type
instead of $d$--wave. We now devote the rest of this section to the precise
statements of these facts.

At any $k\in \lbrack -\pi ,\pi )^{2}$, define the function $\mathbf{s}_{k}:%
\mathbb{Z}^{2}\rightarrow \mathbb{C}$ by%
\begin{equation}
\mathbf{s}_{k}\left( y\right) :=\frac{1}{2}%
\Big (%
\mathrm{e}^{ik\cdot (0,1)}\delta _{y,(0,1)}+\mathrm{e}^{ik\cdot
(0,-1)}\delta _{y,(0,-1)}+\mathrm{e}^{ik\cdot (1,0)}\delta _{y,(1,0)}+%
\mathrm{e}^{ik\cdot (-1,0)}\delta _{y,(-1,0)}%
\Big )
\label{Sk}
\end{equation}%
for all $y\in \mathbb{Z}^{2}$. Let
\begin{equation}
\mathfrak{K}_{v}:=\left\{ k\in \lbrack -\pi ,\pi )^{2}:|\hat{v}%
(k)|=\max_{q\in \lbrack -\pi ,\pi )^{2}}|\hat{v}(q)|=:\Vert \hat{v}\Vert
_{\infty }\right\}  \label{Kv}
\end{equation}%
be the non--empty closed set of maximizers of the absolute value of the
Fourier transform $\hat{v}$ of $v\in \ell ^{1}\left( \mathbb{Z}^{2},\mathbb{R%
}\right) $.

\begin{theorem}[Generic space symmetry of bound pairs]
\label{Carlos super main copy(3)}\mbox{ }\newline
Assume that $\mathfrak{K}_{v}$ is a finite set and take any $\eta >0$. For
sufficiently small $\varepsilon ,\epsilon >0$ and any $(\mathfrak{c}_{t},%
\mathfrak{b}_{t})\in \Xi _{\varepsilon }$, there is a family $%
\{f^{(k)}\}_{k\in \mathfrak{K}}\subset C(\mathbb{R},\ell ^{2}(\mathbb{Z}^{2};%
\mathbb{C}))$ of one--particle wave functions such that:\newline
\emph{(s)} For sufficiently small $U\geq 0$\ and all $t\in \mathbb{R}$,
\begin{equation*}
\sum_{x_{\uparrow },x_{\downarrow }\in \mathbb{Z}^{2}}%
\Big |%
\mathfrak{c}_{t}(x_{\uparrow },x_{\downarrow })-\sum_{k\in \mathfrak{K}%
_{v}}\left\{ \delta _{x_{\uparrow },x_{\downarrow }}+2\mathrm{e}^{-\kappa }%
\mathbf{s}_{k}(x_{\uparrow }-x_{\downarrow })\right\}
f_{t}^{(k)}(x_{\uparrow })%
\Big |%
^{2}\leq \eta \ .
\end{equation*}%
\emph{(d)} For sufficiently large $U>0$\ and all $t\in \mathbb{R}$,
\begin{equation*}
\sum_{x_{\uparrow },x_{\downarrow }\in \mathbb{Z}^{2}}%
\Big |%
\mathfrak{c}_{t}(x_{\uparrow },x_{\downarrow })-\sum_{k\in \mathfrak{K}_{v}}%
\mathbf{s}_{k}(x_{\uparrow }-x_{\downarrow })f_{t}^{(k)}(x_{\uparrow })%
\Big |%
^{2}\leq \eta \ .
\end{equation*}
\end{theorem}

\noindent \textit{Proof.} It is a direct consequence of Corollaries \ref%
{Theorem AC conductivity measure2 copy(2)}--\ref{Theorem AC conductivity
measure2}, Proposition \ref{projector approx} and Lemma \ref{effective BCS
copy(5)}. Note that Proposition \ref{projector approx} only treats the case
of large $U\gg 1$ and Lemma \ref{effective BCS copy(5)} analyzes the $%
\mathfrak{d}$--component of the wave function. To get Assertion (s) we need
similar results for the $\mathfrak{s}$--component at small $U\ll 1$. We omit
the details since the latter case is even simpler.\hfill $\Box $

If the above theorem holds and $\hat{v}$ is concentrated on half breathing
bond-stretching modes, i.e.,
\begin{equation}
\mathfrak{K}_{v}=\{(-\pi ,0),(0,-\pi )\}\subset \lbrack -\pi ,\pi )^{2}
\label{condition sup}
\end{equation}%
(cf. Assumption \ref{MAext}), then the system shows $d$--wave pairing:

\begin{corollary}[$d$--wave space symmetry]
\label{Carlos super main copy(1)}\mbox{ }\newline
Assume (\ref{condition sup}) and take $\eta >0$. For sufficiently small $%
\varepsilon ,\epsilon >0$ and any $(\mathfrak{c}_{t},\mathfrak{b}_{t})\in
\Xi _{\varepsilon }$, there is a one--particle wave function $f_{\varepsilon
}\in C(\mathbb{R},\ell ^{2}(\mathbb{Z}^{2};\mathbb{C}))$ such that, for
sufficiently large $U>0$ and all $t\in \mathbb{R}$,
\begin{equation*}
\sum_{x_{\uparrow },x_{\downarrow }\in \mathbb{Z}^{2}}%
\Big |%
\mathfrak{c}_{t}(x_{\uparrow },x_{\downarrow })-\mathbf{d}(x_{\uparrow
}-x_{\downarrow })f_{t}(x_{\uparrow })%
\Big |%
^{2}\leq \eta \ ,
\end{equation*}%
where
\begin{equation*}
\mathbf{d:=s}_{(-\pi ,0)}=-\mathbf{s}_{(0,-\pi )}.
\end{equation*}
\end{corollary}

If $|\hat{v}(\pm \pi ,0)|$ (or $|\hat{v}(0,\pm \pi )|$) is sufficiently
large, then the wave function of the bound fermion pair has the $d$--wave
symmetry, by Corollary \ref{Carlos super main copy(1)}. Indeed, the Fourier
transform $\mathbf{\hat{d}}$ of $\mathbf{d}$ equals%
\begin{equation*}
\mathbf{\hat{d}}\left( k\right) \equiv \mathbf{\hat{d}}\left(
k_{1},k_{2}\right) =\cos (k_{2})-\cos (k_{1})\text{ }
\end{equation*}%
for any $k\equiv \left( k_{1},k_{2}\right) \in \lbrack -\pi ,\pi )^{2}$.
[See for instance (\ref{fourier transform d wave}).] This is precisely the
orbital function of the $d$--wave pair configuration, see \cite{Bonn}%
.\bigskip

Note that the Hubbard repulsion could be replaced by a general
density--density interaction resulting from the second quantization of
two--body interactions, like for instance (\ref{density-density extension}).
In this case, one has to consider the more general fermion pair annihilation
operator $\tilde{c}_{x}$ as given by (\ref{cooper pair1}) (instead of $c_{x}$%
). Such models would lead to much more general pairing configurations,
beyond $s$-- and $d$--wave orbitals. Basically, if $u\left( r\right) $ has
finite range $[0,R]$ then in the limit $U\rightarrow \infty $ bound fermion
pairs of radius less than $R$ will be suppressed, but the interaction $W$
will bind pairs of fermion separated by a distance of at least $R$, even
when $U\rightarrow \infty $. In this case, the minimum energy of the system
does not depend much on $U$. Similar methods to those used here should be
applicable to such a more general situation. However, we only consider the
most simple physically relevant case $R=0$ to keep technical aspects as
simple as possible.

\section{Uncoupled Effective Models for High--$T_{c}$ Superconductors \label%
{Sedction effective model}}

\subsection{Definition of the Effective Model}

We propose a model which decouples bosons and fermions but which correctly
describes the dynamics ot the orginal model at low energies within the
invariant space $\mathcal{H}_{\uparrow ,\downarrow }^{(2,1)}$, as described
in Section \ref{section main}. The fermionic part is a BCS--like model, as
usually done in theoretical physics, while the bosonic part is a free model
with effective hopping amplitudes.

Indeed, using the decomposition (\ref{decompositino}) we define the bosonic
effective Hamiltonian on the dense subspace $\mathcal{D}$ (\ref{dense
subspace}) by the symmetric operator
\begin{equation}
\tilde{H}_{b}:=-\underset{x,y\in \mathbb{Z}^{2},|x-y|=1}{\sum }w_{b}\left(
x-y\right) b_{x}^{\ast }b_{y}  \label{effective boson model}
\end{equation}%
with%
\begin{equation}
w_{b}\left( x\right) :=\gamma _{b}\left( v_{+}\left( x\right) +v_{-}\left(
x\right) \right) +\frac{1}{2U}\left( v\ast v\right) \left( x\right) \
,\qquad x\in \mathbb{Z}^{2}\ ,  \label{v effecboson}
\end{equation}%
and $\gamma _{b}\geq 0$. Note that the function $w_{b}\in \ell ^{1}\left(
\mathbb{Z}^{2},\mathbb{R}\right) $ is of positive type, and hence the
bosonic hopping amplitudes are of negative type. $w_{b}$ is however not
necessarily positive, as usual hopping terms. Meanwhile, the fermionic
effective Hamiltonian is defined on $\mathcal{D}$ by the symmetric operator%
\begin{eqnarray}
\tilde{H}_{f} &:=&\epsilon \left( -\frac{1}{2}\underset{\mathrm{s}\in
\{\uparrow ,\downarrow \},x,y\in \mathbb{Z}^{2},|x-y|=1}{\sum }a_{x,\mathrm{s}}^{\ast }a_{y,\mathrm{s}}+2\underset{\mathrm{s}\in \{\uparrow ,\downarrow
\},x\in \mathbb{Z}^{2}}{\sum }a_{x,\mathrm{s}}^{\ast }a_{x,\mathrm{s}}\right)   \notag \\
&&+U\underset{x\in \mathbb{Z}^{2}}{\sum }n_{x,\uparrow }n_{x,\downarrow }-\underset{x,y\in \mathbb{Z}^{2}}{\sum }w_{f}\left( x-y\right) c_{x}^{\ast
}c_{y}  \label{effective fermi model}
\end{eqnarray}%
with%
\begin{equation}
w_{f}\left( x\right) :=\gamma _{f}\left( v_{+}\left( x\right) +v_{-}\left(
x\right) \right) -\frac{\gamma _{f}^{2}}{2U+1}\left( v\ast v\right) \left(
x\right) \ ,\qquad x\in \mathbb{Z}^{2}\ ,  \label{v effec}
\end{equation}%
and $\epsilon ,U,\gamma _{f}\geq 0$. See (\ref{cooper pair2}) for the
definition of $c_{x}$. Observe that,\ at large enough $U>0$, the BCS--like
kernel above is of negative type and is thus of \emph{attractive} nature.
The precise form (\ref{v effec}) we have chosen for $w_{f}$ is obtained by
imposing that the effective model gives the exact energy and fermionic
wave--function at orders $U^{0}$ and $U^{-1}$, by expanding this quantities
at any fixed quasi--momentum. Recall that we focus on the large--$U$ regime,
because this is the one related to $d$--wave pairing.

Then, the uncoupled effective model is defined by
\begin{equation}
\mathbf{\tilde{H}}:=\tilde{H}_{f}+\tilde{H}_{b}\in \mathcal{L}(\mathcal{D},%
\mathcal{F}_{-,+})\ .  \label{uncoupled model}
\end{equation}%
Because of (\ref{because of that}), the uncoupled model is 90$%
{{}^\circ}%
$--rotation, reflection and translation invariant, in accordance to
Assumption \ref{MAinv}. \

As in the case of the boson--fermion model $\mathbf{H}$, the subspace $%
\mathcal{H}_{\uparrow ,\downarrow }^{(2,1)}$ is an invariant space of $%
\mathbf{\tilde{H}}$. Therefore, we analyze the dynamics driven by the
bounded self--adjoint operator
\begin{equation*}
\tilde{H}^{(2,1)}:=\overline{\mathbf{\tilde{H}}|_{\mathcal{H}_{\uparrow
,\downarrow }^{(2,1)}}}
\end{equation*}%
at the bottom of its spectrum in order to compare it with the one given by $%
H^{(2,1)}$, see (\ref{important hamiltonian boson fermi}) and (\ref{le chat}%
). The result is the following:

\begin{theorem}[Effectiveness of the uncoupled model]
\label{theorem Gshow}\mbox{ }\newline
Set $\gamma _{b}:=2\mathrm{e}^{-\kappa }$ and $\gamma _{f}=\gamma _{b}^{-1}$%
. Then, there is $\varepsilon _{0}>0$ such that, uniformly for $\varepsilon
\in \left( 0,\varepsilon _{0}\right) $, $t\in \mathbb{R}$, and $\epsilon
,U>0 $,
\begin{equation*}
\left\Vert \left( \mathrm{e}^{-itH^{(2,1)}}-\mathrm{e}^{-it\tilde{H}%
^{(2,1)}}\right) \mathbf{1}_{[E_{0},E_{0}\left( 1-\varepsilon \right)
]}(H^{(2,1)})\right\Vert _{\mathrm{op}}=\mathcal{O}\left( (1+\left\vert
t\right\vert )(\epsilon +U^{-2})\right) \ .
\end{equation*}
\end{theorem}

\noindent \textit{Proof.} The proof is a direct consequence of Lemmata \ref%
{Carlos super main copy(4)} and \ref{effective BCS copy(2)}, Corrollary \ref%
{Theorem AC conductivity measure2}, Equations (\ref{fiber decomposition})--(%
\ref{fiber decomposition2}), and Propositions \ref{pairing mode copy(1)}--%
\ref{proposition Self--adjoint decomposable operators}.\hfill $\Box $

If the functions $v_{\pm }$ decay sufficiently fast in space, then we can
find kernels $w_{b},w_{f}\in \ell ^{1}\left( \mathbb{Z}^{2},\mathbb{R}%
\right) $ of positive type such that
\begin{equation*}
\left\Vert \left( \mathrm{e}^{-itH^{(2,1)}}-\mathrm{e}^{-it\tilde{H}%
^{(2,1)}}\right) \mathbf{1}_{[E_{0},E_{0}\left( 1-\varepsilon \right)
]}(H^{(2,1)})\right\Vert _{\mathrm{op}}=\mathcal{O}\left( \epsilon
+U^{-2}\right) \ ,
\end{equation*}%
uniformly in time $t\in \mathbb{R}$. Indeed, when $E_{0}<0$, one chooses $%
w_{f}$ and $w_{b}$ with Fourier transform $\hat{w}_{f}$ and $\hat{w}_{b}$,
respectively, such that
\begin{equation*}
\hat{w}_{f}\left( k\right) \mathcal{R}(k,U,\mathrm{E}(k))=1
\end{equation*}%
and $\hat{w}_{b}(k)=\mathrm{E}(k)$ for $k$ in open neighborhood of the set
of minimizers of $\mathrm{E}(\cdot )$ on $[-\pi ,\pi )^{2}$. See Proposition %
\ref{effective BCS copy(1)}, Theorem \ref{effective BCS copy(3)}, and
Equation (\ref{estimate}).\

\subsection{Long--Range Idealization of the Effective Electron--Electron
Interaction\label{Section effec1}}

The analysis of equilibrium states of fermionic models like (\ref{effective
fermi model}) is known to be a very difficult task. Indeed, the complete
phase diagram of the Hubbard model, which is (\ref{effective fermi model})
with $w_{f}\equiv 0$, is still unknown for dimensions bigger than one, at
least in a mathematically rigorous sense. However, for certain classes of
long--range couplings $w_{f}$, the equilibrium states of (\ref{effective
fermi model}) become much more tractable: We showed in \cite{BruPedra2} how
to construct equilibrium states of long--range models as convex combinations
of equilibrium states of much more simple \textquotedblleft Bogolioubov
approximations\textquotedblright\ of the starting model.

By Assumption \ref{MAext}, the Fourier transform $\hat{v}$ of the coupling
function $v$ is concentrated around a few points in the Brillouin zone. See (%
\ref{condition sup}). This implies from (\ref{v effec}) that $\hat{w}_{f}$
is also concentrated around the same points. We can thus consider the
idealization where the Fourier transform $\hat{w}_{f}$ tends to a
distribution supported on that few points. More precisely, it is reasonable
to replace $\hat{w}_{f}$ by
\begin{equation*}
w_{f}^{(\mathrm{MF})}\left( x\right) :=\gamma _{f}^{(\mathrm{MF})}\left(
\mathrm{e}^{i(-\pi ,0)\cdot x}+\mathrm{e}^{i(0,-\pi )\cdot x}\right)
\end{equation*}%
for $x\in \mathbb{Z}^{2}$, where $\gamma _{f}^{(\mathrm{MF})}\geq 0$ is a
positive constant. However, this kernel is not anymore summable in the $%
\mathbb{Z}^{2}$--lattice. The corresponding interaction\ has thus to be
interpreted as a mean field term. Hence, we define the corresponding mean
field type model in cubic boxes $\Lambda _{l}\subset \mathbb{Z}^{2}$ of size
length $l\in \mathbb{R}^{+}$ with volume $\left\vert \Lambda _{l}\right\vert
$:%
\begin{eqnarray}
\tilde{H}_{f,l}^{(\mathrm{MF})} &:=&\epsilon \left( -\frac{1}{2}\underset{\mathrm{s}\in \{\uparrow ,\downarrow \},x,y\in \Lambda _{l},|x-y|=1}{\sum }a_{x,\mathrm{s}}^{\ast }a_{y,\mathrm{s}}+2\underset{\mathrm{s}\in \{\uparrow
,\downarrow \},x\in \Lambda _{l}}{\sum }a_{x,\mathrm{s}}^{\ast }a_{x,\mathrm{s}}\right)   \notag \\
&&+2U\underset{x\in \Lambda _{l}}{\sum }n_{x,\uparrow }n_{x,\downarrow }-\frac{1}{\left\vert \Lambda _{l}\right\vert }\underset{x,y\in \Lambda _{l}}{\sum }w_{f}^{(\mathrm{MF})}\left( x-y\right) c_{x}^{\ast }c_{y}\ .
\label{Gshow Hamil}
\end{eqnarray}%
Compare with (\ref{effective fermi model}). Observe that the last term is $%
\left\vert \Lambda _{l}\right\vert $ times the sum of the squares of the
space averages of two operators:%
\begin{eqnarray*}
&&\frac{1}{\left\vert \Lambda _{l}\right\vert }\underset{x,y\in \Lambda _{l}}%
{\sum }w_{f}^{(\mathrm{MF})}\left( x-y\right) c_{x}^{\ast }c_{y} \\
&=&\gamma ^{(\mathrm{MF})}\left\vert \Lambda _{l}\right\vert \left( \frac{1}{%
\left\vert \Lambda _{l}\right\vert }\underset{x\in \Lambda _{l}}{\sum }%
\mathrm{e}^{i(\pi ,0)\cdot x}c_{x}\right) ^{\ast }\left( \frac{1}{\left\vert
\Lambda _{l}\right\vert }\underset{x\in \Lambda _{l}}{\sum }\mathrm{e}%
^{i(\pi ,0)\cdot x}c_{x}\right) \\
&&+\gamma ^{(\mathrm{MF})}\left\vert \Lambda _{l}\right\vert \left( \frac{1}{%
\left\vert \Lambda _{l}\right\vert }\underset{x\in \Lambda _{l}}{\sum }%
\mathrm{e}^{i(0,\pi )\cdot x}c_{x}\right) ^{\ast }\left( \frac{1}{\left\vert
\Lambda _{l}\right\vert }\underset{x\in \Lambda _{l}}{\sum }\mathrm{e}%
^{i(0,\pi )\cdot x}c_{x}\right) \ .
\end{eqnarray*}%
This leads to a long--range interaction (cf. Assumption \ref{MAlr}) similar
to the ones treated in \cite{BruPedra2}. The long--range component of the
model discussed here is rather a sum over periodic (but not translation
invariant) mean field type quadratic terms. The methods of \cite{BruPedra2}
have to be adapted to this case, but they are still applicable.

In this case, one has to be able to study the \textquotedblleft Bogolioubov
approximations\textquotedblright\ of the model $\tilde{H}_{f,l}^{(\mathrm{MF}%
)}$, at least in the strong coupling regime (Cf. Assumption \ref{MArep
copy(1)}), i.e., for $\epsilon =0$. It is also important to check that the
behavior of the system is not singular at $\epsilon =0$. In this context,
methods of constructive quantum field theory, as Grassmann--Berezin
integrations, Brydges--Kennedy tree expansions and determinant bounds \cite%
{Pedra-Salmhofer} will be important. We recently applied such methods in a
similar situation in \cite{meissner} to analyze the Meissner--Ochsenfeld
effect, starting from a microscopic model. Technically speaking, this last
study is difficult. We plan to work out these problems in subsequent papers.

\section{Technical Proofs\label{interesting sect proofs}}

Before starting, note that the computations are given in all details to make
them \textquotedblleft self-contained\textquotedblright\ and hence
accessible to readers not used with the methods. Even if it is not
explicitly mentioned, we always have $\epsilon ,U,h_{b}\geq 0$ and $\kappa >0
$. For simplicity and without loss of generality, in this section we
sometimes fix $h_{b}\in \left[ 0,1\right] $.

\subsection{Fiber Decomposition of the 2-Fermions--1-Boson Hamiltonian}

The (fermionic and bosonic) kinetic parts of the Hamiltonian $H^{(2,1)}$
defined by (\ref{important hamiltonian boson fermi}) are diagonalizable by
the Fourier transform. The interaction term (\ref{interaction W}) is such
that it annihilates either a boson to create a fermion pair or a fermion
pair to create a boson with same total quasi--momentum, in both cases. As a
consequence, it is natural to decompose $H^{(2,1)}$ on fibers parametrized
by Fourier modes $k\in \lbrack -\pi ,\pi )^{2}$, which stand for total
quasi--momenta on the torus. It is done as follows:

We denote the Haar measure on the torus $[-\pi ,\pi )^{2}$ by $\mathfrak{m}$%
, i.e.,
\begin{equation*}
\mathfrak{m}\left( \mathrm{d}^{2}q\right) :=(2\pi )^{-2}\mathrm{d}^{2}q\ .
\end{equation*}%
Using the direct integral of Hilbert spaces (see Section \ref{sect direct
decomposition}), let%
\begin{eqnarray*}
\mathfrak{F}_{\uparrow ,\downarrow }^{(2,1)} &:=&\int_{[-\pi ,\pi
)^{2}}^{\oplus }L^{2}([-\pi ,\pi )^{2},\mathfrak{m};\mathbb{C})\times
\mathbb{C}\ \mathfrak{m}(\mathrm{d}^{2}k) \\
&\simeq &\int_{[-\pi ,\pi )^{2}}^{\oplus }L^{2}([-\pi ,\pi )^{2},\mathfrak{m};\mathbb{C})\ \mathfrak{m}(\mathrm{d}^{2}k)\times L^{2}([-\pi ,\pi )^{2},\mathfrak{m};\mathbb{C}) \ .
\end{eqnarray*}%
This space is also unitarily equivalent to the Hilbert space%
\begin{equation*}
\mathcal{H}_{\uparrow ,\downarrow }^{(2,1)}\simeq \ell ^{2}(\mathbb{Z}%
^{2}\times \mathbb{Z}^{2};\mathbb{C})\times \ell ^{2}(\mathbb{Z}^{2};\mathbb{%
C})\ ,
\end{equation*}%
see (\ref{hilbert space elementaire1}) and (\ref{hilbert space elementaire2}%
). An isometry between both spaces is defined by
\begin{equation}
\mathfrak{U}(\mathfrak{\hat{c}},\mathfrak{\hat{b}}):=(\mathfrak{U}_{\uparrow
\downarrow }(\mathfrak{\hat{c}}),\mathfrak{U}_{b}(\mathfrak{\hat{b}}))\in
\mathcal{H}_{\uparrow ,\downarrow }^{(2,1)}\ ,\quad (\mathfrak{\hat{c}},%
\mathfrak{\hat{b}})\in \mathfrak{F}_{\uparrow ,\downarrow }^{(2,1)}\ ,
\label{defnition unitary1}
\end{equation}%
where the wave function $\mathfrak{U}_{\uparrow \downarrow }(\mathfrak{\hat{c%
}})$ of one fermion pair in $\mathcal{H}_{\uparrow ,\downarrow }^{(2,1)}$
equals%
\begin{eqnarray}
\lbrack \mathfrak{U}_{\uparrow \downarrow }(\mathfrak{\hat{c}})](x_{\uparrow
},x_{\downarrow }) &:=&\int_{[-\pi ,\pi )^{2}}\mathfrak{m}(\mathrm{d}^{2}k)\
\int_{[-\pi ,\pi )^{2}}\mathfrak{m}(\mathrm{d}^{2}k_{\uparrow \downarrow })
\notag \\
&&\quad \mathrm{e}^{ik\cdot x_{\uparrow }}\mathrm{e}^{ik_{\uparrow
\downarrow }\cdot (x_{\downarrow }-x_{\uparrow })}\left[ \mathfrak{\hat{c}}(k)\right] (k_{\uparrow \downarrow })\ ,  \label{defnition unitary2}
\end{eqnarray}%
for any $x_{\uparrow },x_{\downarrow }\in \mathbb{Z}^{2}$, while the wave
function $\mathfrak{U}_{b}(\mathfrak{\hat{b}})$ of one boson is
\begin{equation}
\lbrack \mathfrak{U}_{b}(\mathfrak{\hat{b}})](x_{b}):=\int_{[-\pi ,\pi )^{2}}%
\mathrm{e}^{ik\cdot x_{b}}\mathfrak{\hat{b}}(k)\ \mathfrak{m}(\mathrm{d}%
^{2}k)  \label{defnition unitary3}
\end{equation}%
for $x_{b}\in \mathbb{Z}^{2}$. Since
\begin{equation}
L^{2}([-\pi ,\pi )^{2}\times \lbrack -\pi ,\pi )^{2},\mathfrak{m}\otimes
\mathfrak{m};\mathbb{C})\subset L^{1}([-\pi ,\pi )^{2}\times \lbrack -\pi
,\pi )^{2},\mathfrak{m}\otimes \mathfrak{m};\mathbb{C})\ ,
\label{Gshwo idiot}
\end{equation}%
and\
\begin{equation}
L^{2}([-\pi ,\pi )^{2},\mathfrak{m};\mathbb{C})\subset L^{1}([-\pi ,\pi
)^{2},\mathfrak{m};\mathbb{C})\ ,  \label{Gshwo idiot2}
\end{equation}%
note that the r.h.s. of (\ref{defnition unitary2})--(\ref{defnition unitary3}%
) are well--defined. Moreover, the operators
\begin{equation*}
\mathfrak{U}_{\uparrow \downarrow }:\int_{[-\pi ,\pi )^{2}}^{\oplus
}L^{2}([-\pi ,\pi )^{2},\mathfrak{m};\mathbb{C})\ \mathfrak{m}(\mathrm{d}%
^{2}k)\rightarrow \ell ^{2}(\mathbb{Z}^{2}\times \mathbb{Z}^{2};\mathbb{C})
\end{equation*}%
and%
\begin{equation*}
\mathfrak{U}_{b}:L^{2}([-\pi ,\pi )^{2},\mathfrak{m};\mathbb{C})\rightarrow
\ell ^{2}(\mathbb{Z}^{2};\mathbb{C})
\end{equation*}%
are unitary. The inverse $\mathfrak{U}^{-1}=\mathfrak{U}^{\ast }$ is%
\begin{equation}
\mathfrak{U}^{\ast }=\mathfrak{U}_{\uparrow \downarrow }^{\ast }\oplus
\mathfrak{U}_{b}^{\ast }  \label{u stra}
\end{equation}%
and $\mathfrak{F}_{\uparrow ,\downarrow }^{(2,1)}=\mathfrak{U}^{\ast }%
\mathcal{H}_{\uparrow ,\downarrow }^{(2,1)}$ (using (\ref{hilbert space
elementaire2})).

To obtain explicit expressions for the actions of $\mathfrak{U}_{\uparrow
\downarrow }^{\ast }$ and $\mathfrak{U}_{b}^{\ast }$, it suffices to
consider dense subspaces of $\ell ^{2}(\mathbb{Z}^{2}\times \mathbb{Z}^{2};%
\mathbb{C})$ and $\ell ^{2}(\mathbb{Z}^{2};\mathbb{C})$, respectively. Note
indeed that, in contrast with\ (\ref{Gshwo idiot})--(\ref{Gshwo idiot2}),
\begin{equation*}
\ell ^{1}(\mathbb{Z}^{2}\times \mathbb{Z}^{2};\mathbb{C})\varsubsetneq \ell
^{2}(\mathbb{Z}^{2}\times \mathbb{Z}^{2};\mathbb{C})\text{\quad and\quad }%
\ell ^{1}(\mathbb{Z}^{2};\mathbb{C})\varsubsetneq \ell ^{2}(\mathbb{Z}^{2};%
\mathbb{C})\ .
\end{equation*}%
For any $\mathfrak{c}\in \ell ^{1}(\mathbb{Z}^{2}\times \mathbb{Z}^{2};%
\mathbb{C})$ and $k,k_{\uparrow \downarrow }\in \lbrack -\pi ,\pi )^{2}$,
\begin{equation*}
\left[ \mathfrak{U}_{\uparrow \downarrow }^{\ast }(\mathfrak{c})(k)\right]
(k_{\uparrow \downarrow })=\sum_{x_{\uparrow },x_{\uparrow \downarrow }\in
\mathbb{Z}^{2}}\mathrm{e}^{-ik\cdot x_{\uparrow }}\mathrm{e}^{-ik_{\uparrow
\downarrow }\cdot x_{\uparrow \downarrow }}\mathfrak{c}(x_{\uparrow
},x_{\uparrow }+x_{\uparrow \downarrow })\ ,
\end{equation*}%
while, for any $\mathfrak{b}\in \ell ^{1}(\mathbb{Z}^{2};\mathbb{C})$ and $%
k\in \lbrack -\pi ,\pi )^{2}$,
\begin{equation*}
\mathfrak{U}_{b}^{\ast }(\mathfrak{b})(k)=\sum_{x_{b}\in \mathbb{Z}^{2}}%
\mathrm{e}^{-ik\cdot x_{b}}\mathfrak{b}(x_{b})\ .
\end{equation*}

Now we study the operator $\mathfrak{U}^{\ast }H^{(2,1)}\mathfrak{U}$ acting
on $\mathfrak{F}_{\uparrow ,\downarrow }^{(2,1)}$ (cf. (\ref{important
hamiltonian boson fermi})). We start by deriving that its fiber
decomposition (see Section \ref{sect direct decomposition} for more
details). To this end, recall that $\hat{v}$ is the Fourier transform of the
coupling function $v$ and, at each $k\in \lbrack -\pi ,\pi )^{2}$, let $%
\mathfrak{s},\mathfrak{d}(k)\in L^{2}([-\pi ,\pi )^{2},\mathfrak{m};\mathbb{C%
})$ be defined by%
\begin{equation}
\mathfrak{s}(k_{\uparrow \downarrow }):=1\text{\qquad and\qquad }[\mathfrak{d%
}(k)](k_{\uparrow \downarrow }):=2\mathrm{e}^{-\kappa }\cos (k_{\uparrow
\downarrow }-k)\ ,  \label{def d et s}
\end{equation}%
for all $k_{\uparrow \downarrow }\in \lbrack -\pi ,\pi )^{2}$, where the
function $\cos $ is defined on the torus $[-\pi ,\pi )^{2}$ by (\ref{cosinus}%
). Let $P_{\mathfrak{d}(k)}=P_{\mathfrak{d}(k)}^{\ast }\in \mathcal{B}%
(L^{2}([-\pi ,\pi )^{2},\mathfrak{m};\mathbb{C}))$ be the orthogonal
projection with range%
\begin{equation}
\mathrm{Ran}(P_{\mathfrak{d}(k)})=\mathbb{C(}\mathfrak{d}(k)+\mathfrak{s})\ .
\label{range}
\end{equation}%
Similarly, $P_{0}=P_{0}^{\ast }\in \mathcal{B}(L^{2}([-\pi ,\pi )^{2},%
\mathfrak{m};\mathbb{C}))$ is the orthogonal projection with range%
\begin{equation}
\mathrm{Ran}(P_{0})=\mathbb{C}\mathfrak{s}\ .  \label{range0}
\end{equation}%
At any $k\in \lbrack -\pi ,\pi )^{2}$, define now the bounded operators
\begin{eqnarray*}
A_{1,1}^{(0)}(k) &:&L^{2}([-\pi ,\pi )^{2},\mathfrak{m};\mathbb{C}%
)\rightarrow L^{2}([-\pi ,\pi )^{2},\mathfrak{m};\mathbb{C})\ , \\
A_{1,1}(k) &:&L^{2}([-\pi ,\pi )^{2},\mathfrak{m};\mathbb{C})\rightarrow
L^{2}([-\pi ,\pi )^{2},\mathfrak{m};\mathbb{C})\ , \\
A_{2,1}(k) &:&L^{2}([-\pi ,\pi )^{2},\mathfrak{m};\mathbb{C})\rightarrow
\mathbb{C}\ , \\
A_{1,2}(k) &:&\mathbb{C}\rightarrow L^{2}([-\pi ,\pi )^{2},\mathfrak{m};%
\mathbb{C})\ , \\
A_{2,2}(k) &:&\mathbb{C}\rightarrow \mathbb{C}\ ,
\end{eqnarray*}%
by%
\begin{eqnarray}
\lbrack A_{1,1}^{(0)}(k)\Psi _{\uparrow \downarrow }](k_{\uparrow \downarrow
}) &:=&\epsilon \left( 4-\cos (k_{\uparrow \downarrow }-k)-\cos (k_{\uparrow
\downarrow })\right) \Psi _{\uparrow \downarrow }(k_{\uparrow \downarrow })\
,  \label{A legallll1} \\
A_{1,1}(k)\Psi _{\uparrow \downarrow } &:=&A_{1,1}^{(0)}(k)\Psi _{\uparrow
\downarrow }+UP_{0}\Psi _{\uparrow \downarrow }\ ,  \label{A legall0} \\
A_{2,1}(k)\Psi _{\uparrow \downarrow } &:=&\hat{v}\left( k\right)
\left\langle \mathfrak{d}(k)+\mathfrak{s},\Psi _{\uparrow \downarrow
}\right\rangle \ ,  \label{A legallll2} \\
A_{1,2}(k)\Psi _{b} &:=&\Psi _{b}\hat{v}\left( k\right) \left( \mathfrak{d}(k)+\mathfrak{s}\right) \ ,  \label{A legallll3} \\
A_{2,2}(k)\Psi _{b} &:=&\epsilon h_{b}(2-\cos (k))\Psi _{b}\ ,
\label{A legallll4}
\end{eqnarray}
for all $\Psi _{\uparrow \downarrow }\in L^{2}([-\pi ,\pi )^{2},\mathfrak{m};%
\mathbb{C})$ and $\Psi _{b}\in $ $\mathbb{C}$. Here, $\left\langle \cdot
,\cdot \right\rangle $ stands for the scalar product of $L^{2}([-\pi ,\pi
)^{2},\mathfrak{m};\mathbb{C})$. By continuity of $\hat{v}$, the maps $%
k\mapsto A_{i,j}(k)$ are continuous, in operator norm sense, for all $i,j\in
\{1,2\}$. In particular,%
\begin{equation}
A(\cdot ):=\left(
\begin{array}{cc}
A_{1,1}(\cdot ) & A_{1,2}(\cdot ) \\
A_{2,1}(\cdot ) & A_{2,2}(\cdot )%
\end{array}%
\right) \in L^{\infty }([-\pi ,\pi )^{2},\mathfrak{m};\mathcal{B}%
(L^{2}([-\pi ,\pi )^{2},\mathfrak{m};\mathbb{C})\times \mathbb{C}))\ .
\label{A legallll5}
\end{equation}%
By \cite[Theorem XIII.83]{RS4} (see also Section \ref{sect direct
decomposition}), there is a unique decomposable operator%
\begin{equation}
A:=\int_{[-\pi ,\pi )^{2}}^{\oplus }A(k)\text{ }\mathfrak{m}(\mathrm{d}%
^{2}k)\in \mathcal{B}(\mathfrak{F}_{\uparrow ,\downarrow }^{(2,1)})\ ,
\label{A legallll}
\end{equation}%
which turns out to coincide with $\mathfrak{U}^{\ast }H^{(2,1)}\mathfrak{U}$:

\begin{lemma}[Direct integral decomposition]
\label{Carlos super main copy(4)}%
\begin{equation*}
A=\mathfrak{U}^{\ast }H^{(2,1)}\mathfrak{U}\qquad \text{and}\qquad \Vert
H^{(2,1)}\Vert _{\mathrm{op}}=\max_{k\in \lbrack -\pi ,\pi )^{2}}\Vert
A(k)\Vert _{\mathrm{op}}\ .
\end{equation*}
\end{lemma}

\noindent \textit{Proof.} Define the dense set
\begin{equation*}
\mathcal{D}_{\uparrow ,\downarrow }^{(2,1)}:=\mathfrak{U}^{\ast }\left[ \ell
^{1}(\mathbb{Z}^{2}\times \mathbb{Z}^{2};\mathbb{C})\times \ell ^{1}(\mathbb{%
Z}^{2};\mathbb{C})\right] \subset \mathfrak{F}_{\uparrow ,\downarrow
}^{(2,1)}\ .
\end{equation*}%
For any $(\mathfrak{\hat{c}},\mathfrak{\hat{b}})\in \mathcal{D}_{\uparrow
,\downarrow }^{(2,1)}$, we infer from (\ref{interesting hamiltonian free
electrons}), (\ref{boson hamiltonian}), (\ref{interaction W0}), (\ref%
{interaction W}), (\ref{cooper pair2}), and (\ref{idiot}) that%
\begin{equation*}
H^{(2,1)}\mathfrak{U}(\mathfrak{\hat{c}},\mathfrak{\hat{b}})=(\mathfrak{c}%
^{\prime },\mathfrak{b}^{\prime })\in \mathcal{H}_{\uparrow ,\downarrow
}^{(2,1)}\ ,
\end{equation*}%
where, for any $x_{\uparrow },x_{\downarrow }\in \mathbb{Z}^{2}$,%
\begin{eqnarray*}
\mathfrak{c}^{\prime }(x_{\uparrow },x_{\downarrow }) &=&-\frac{\epsilon }{2}%
\underset{z\in \mathbb{Z}^{2},\text{ }|z|=1}{\sum }[\mathfrak{U}_{\uparrow
\downarrow }(\mathfrak{\hat{c}})](x_{\uparrow }+z,x_{\downarrow })-\frac{%
\epsilon }{2}\underset{z\in \mathbb{Z}^{2},\text{ }|z|=1}{\sum }[\mathfrak{U}%
_{\uparrow \downarrow }(\mathfrak{\hat{c}})](x_{\uparrow },x_{\downarrow }+z)
\\
&&+4\epsilon \lbrack \mathfrak{U}_{\uparrow \downarrow }(\mathfrak{\hat{c}}%
)](x_{\uparrow },x_{\downarrow })+U\delta _{x_{\uparrow },x_{\downarrow }}[%
\mathfrak{U}_{\uparrow \downarrow }(\mathfrak{\hat{c}})](x_{\uparrow
},x_{\downarrow }) \\
&&+\underset{x_{b}\in \mathbb{Z}^{2}}{\sum }v\left( x_{\downarrow
}-x_{b}\right) \delta _{x_{\uparrow },x_{\downarrow }}[\mathfrak{U}_{b}(%
\mathfrak{\hat{b}})](x_{b}) \\
&&+\mathrm{e}^{-\kappa }\underset{x_{b}\in \mathbb{Z}^{2}}{\sum }\ \underset{%
z\in \mathbb{Z}^{2},\text{ }|z|=1}{\sum }v\left( x_{\downarrow
}-x_{b}\right) \delta _{x_{\uparrow }+z,x_{\downarrow }}[\mathfrak{U}_{b}(%
\mathfrak{\hat{b}})](x_{b})
\end{eqnarray*}%
and, for any $x_{b}\in \mathbb{Z}^{2}$,
\begin{eqnarray*}
\mathfrak{b}^{\prime }(x_{b}) &=&-\frac{\epsilon h_{b}}{2}\underset{z\in
\mathbb{Z}^{2},\text{ }|z|=1}{\sum }[\mathfrak{U}_{b}(\mathfrak{\hat{b}}%
)](x_{b}+z)+2\epsilon h_{b}[\mathfrak{U}_{b}(\mathfrak{\hat{b}})](x_{b}) \\
&&+\sum_{x\in \mathbb{Z}^{2}}v\left( x_{b}-x\right) [\mathfrak{U}_{\uparrow
\downarrow }(\mathfrak{\hat{c}})](x,x) \\
&&+\mathrm{e}^{-\kappa }\sum_{x\in \mathbb{Z}^{2}}\underset{z\in \mathbb{Z}%
^{2},\text{ }|z|=1}{\sum }v\left( x_{b}-x\right) [\mathfrak{U}_{\uparrow
\downarrow }(\mathfrak{\hat{c}})](x+z,x)\ .
\end{eqnarray*}%
Therefore, for any $(\mathfrak{\hat{c}},\mathfrak{\hat{b}})\in \mathcal{D}%
_{\uparrow ,\downarrow }^{(2,1)}$,
\begin{equation*}
\mathfrak{U}^{\ast }H^{(2,1)}\mathfrak{U}(\mathfrak{\hat{c}},\mathfrak{\hat{b%
}})=(\mathfrak{\hat{c}}^{\prime },\mathfrak{\hat{b}}^{\prime })\in \mathfrak{%
F}_{\uparrow ,\downarrow }^{(2,1)}\ ,
\end{equation*}%
where, for any $k,k_{\uparrow \downarrow }\in \lbrack -\pi ,\pi )^{2}$,
\begin{eqnarray*}
\lbrack \mathfrak{\hat{c}}^{\prime }(k)](k_{\uparrow \downarrow })
&=&\epsilon \left( 4-\cos (k_{\uparrow \downarrow }-k)-\cos (k_{\uparrow
\downarrow })\right) [\mathfrak{\hat{c}}(k)](k_{\uparrow \downarrow }) \\
&&+U\int_{[-\pi ,\pi )^{2}}[\mathfrak{\hat{c}}(k)](k_{\uparrow \downarrow
})\ \mathfrak{m}(\mathrm{d}^{2}k_{\uparrow \downarrow }) \\
&&+\hat{v}\left( k\right) (1+2\mathrm{e}^{-\kappa }\cos (k_{\uparrow
\downarrow }-k))\mathfrak{\hat{b}}(k)\ , \\
\mathfrak{\hat{b}}^{\prime }(k) &=&\epsilon h_{b}(2-\cos (k))\mathfrak{\hat{b%
}}(k) \\
&&+\hat{v}\left( k\right) \int_{[-\pi ,\pi )^{2}}(1+2\mathrm{e}^{-\kappa
}\cos (k_{\uparrow \downarrow }-k))\left[ \mathfrak{\hat{c}}(k)\right]
(k_{\uparrow \downarrow })\ \mathfrak{m}(\mathrm{d}^{2}k_{\uparrow
\downarrow })\ .
\end{eqnarray*}%
By (\ref{A legallll1})--(\ref{A legallll}), it follows that $\mathfrak{U}%
^{\ast }H^{(2,1)}\mathfrak{U}=A$ on the dense subspace $\mathcal{D}%
_{\uparrow ,\downarrow }^{(2,1)}$. By the boundedness of both operators on $%
\mathfrak{F}_{\uparrow ,\downarrow }^{(2,1)}$, we arrive at the first
assertion. To prove that%
\begin{equation*}
\Vert A\Vert _{\mathrm{op}}=\max_{k\in \lbrack -\pi ,\pi )^{2}}\Vert
A(k)\Vert _{\mathrm{op}}\text{ },
\end{equation*}%
and thus the second assertion, note that%
\begin{equation*}
\Vert A\Vert _{\mathrm{op}}=\underset{k\in \lbrack -\pi ,\pi )^{2}}{\mathrm{%
ess}\text{ }\sup }\Vert A(k)\Vert _{\mathrm{op}}\text{ }.
\end{equation*}%
See Section \ref{sect direct decomposition}. Now, to complete the proof, use
the continuity of the map $k\mapsto A(k)$. \hfill $\Box $

By using Lemma \ref{Carlos super main copy(4)} and Proposition \ref%
{proposition Self--adjoint decomposable operators}, we can extract spectral
properties of $H^{(2,1)}$. In particular, the spectrum $\sigma (H^{(2,1)})$
of $H^{(2,1)}$ is bounded from below by
\begin{equation}
E_{0}:=\inf \sigma (H^{(2,1)})\geq \inf_{k\in \lbrack -\pi ,\pi
)^{2}}\left\{ \min \sigma (A(k))\right\} \ .  \label{ess spectrum0}
\end{equation}%
In fact, the latter bound holds with equality:

\begin{lemma}[Bottom of the spectrum of $H^{(2,1)}$]
\label{Carlos super main copy(5)}\mbox{ }\newline
For any $k\in \lbrack -\pi ,\pi )^{2}$, let $\sigma _{\mathrm{d}}(A(k))$ be
the discrete spectrum of $A(k)$. Then%
\begin{eqnarray*}
E_{0} &=&\min \sigma (H^{(2,1)})=\min_{k\in \lbrack -\pi ,\pi )^{2}}\left\{
\min \sigma (A(k))\right\} \\
&=&\min \left\{ 0,\min_{k\in \lbrack -\pi ,\pi )^{2}}\left\{ \min \sigma _{%
\mathrm{d}}(A(k))\right\} \right\} \leq 0\ .
\end{eqnarray*}
\end{lemma}

\noindent \textit{Proof.} The operators $A_{2,1}$, $A_{1,2}$ and $P_{0}$ are
compact operators. Hence, for any $k\in \lbrack -\pi ,\pi )^{2}$, the
essential spectrum $\sigma _{\mathrm{ess}}(A(k))$ of $A(k)$ equals%
\begin{equation}
\sigma _{\mathrm{ess}}(A(k))=2\epsilon \cos (k/2)\cdot \lbrack
-1,1]+4\epsilon \subset \left[ 0,8\epsilon \right] \text{\ }
\label{ess spectrum}
\end{equation}%
for any $\epsilon >0$, while $\sigma _{\mathrm{ess}}(A(k))=\emptyset $ when $%
\epsilon =0$. It follows from (\ref{ess spectrum0})--(\ref{ess spectrum})
and Proposition \ref{proposition Self--adjoint decomposable operators}
together with Kato's theory for the perturbation of the discrete spectrum $%
\sigma _{\mathrm{d}}$ of closed operators that%
\begin{equation}
E_{0}=\min \sigma (H^{(2,1)})=\min_{k\in \lbrack -\pi ,\pi )^{2}}\left\{
\min \left\{ \sigma _{\mathrm{ess}}(A(k))\cup \sigma _{\mathrm{d}%
}(A(k))\right\} \right\} \ .  \label{spectrum bis}
\end{equation}%
Since
\begin{equation}
\min_{k\in \lbrack -\pi ,\pi )^{2}}\left\{ \min \sigma _{\mathrm{ess}%
}(A(k))\right\} =0\ ,  \label{spectrum bisbis}
\end{equation}%
we thus infer the assertion from (\ref{spectrum bis}).\hfill $\Box $

\subsection{Negative Eigenvalues of the Fiber Hamiltonians}

We analyze now the bottom of the spectrum of the fiber Hamiltonians $A(k)$
for quasi--momenta $k\in \lbrack -\pi ,\pi )^{2}$:

\begin{lemma}[Negative eigenvalues of $A(k)$ -- I]
\label{effective BCS}\mbox{ }\newline
Let $k\in \lbrack -\pi ,\pi )^{2}$ and $\lambda <0$. Then, $\lambda \in
\sigma (A(k))$ if and only if $v\left( k\right) \neq 0$ and there is $\Psi
_{\uparrow \downarrow }\in L^{2}([-\pi ,\pi )^{2},\mathfrak{m};\mathbb{C}%
)\backslash \{0\}$ such that%
\begin{equation*}
\left[ (A_{1,1}(k)-\lambda )-A_{1,2}(k)(A_{2,2}(k)-\lambda )^{-1}A_{2,1}(k)%
\right] \Psi _{\uparrow \downarrow }=0\ .
\end{equation*}%
In this case, $\lambda <0$ is an eigenvalue of $A(k)$ with associated
eigenvector
\begin{equation*}
\left( \Psi _{\uparrow \downarrow },-(A_{2,2}(k)-\lambda
)^{-1}A_{2,1}(k)\Psi _{\uparrow \downarrow }\right) \in L^{2}([-\pi ,\pi
)^{2},\mathfrak{m};\mathbb{C})\backslash \{0\}\times \mathbb{C}\backslash
\{0\}\ .
\end{equation*}
\end{lemma}

\noindent \textit{Proof.} Fix $k\in \lbrack -\pi ,\pi )^{2}$ and $\lambda <0$%
. Assume that $\lambda \in \sigma (A(k))$. Then $\lambda \in \sigma _{%
\mathrm{d}}(A(k))$, by (\ref{ess spectrum}). For such a discrete eigenvalue
there is $\Psi _{\uparrow \downarrow }\in L^{2}([-\pi ,\pi )^{2},\mathfrak{m}%
;\mathbb{C})$ and $\Psi _{b}\in \mathbb{C}$ such that $(\Psi _{\uparrow
\downarrow },\Psi _{b})\neq (0,0)$ and%
\begin{equation}
\left(
\begin{array}{l}
(A_{1,1}(k)-\lambda )\Psi _{\uparrow \downarrow }+A_{1,2}(k)\Psi _{b} \\
A_{2,1}(k)\Psi _{\uparrow \downarrow }+(A_{2,2}(k)-\lambda )\Psi _{b}%
\end{array}%
\right) =\left(
\begin{array}{l}
0 \\
0%
\end{array}%
\right) \ ,  \label{eq super macarena}
\end{equation}%
see (\ref{A legallll5}). From (\ref{A legallll1})--(\ref{A legall0}) and (%
\ref{A legallll4}) with $U,\epsilon ,h_{b}\geq 0$ and $\lambda <0$, note
that $A_{2,2}(k)-\lambda >0$ and $A_{1,1}(k)-\lambda >0$. This yields $\Psi
_{\uparrow \downarrow }\neq 0$, $\Psi _{b}\neq 0$ and $v\left( k\right) \neq
0$. Hence, if $\lambda <0$ then $\lambda \in \sigma (A(k))$ if and only if (%
\ref{eq super macarena}) holds true with $\Psi _{\uparrow \downarrow }\in
L^{2}([-\pi ,\pi )^{2},\mathfrak{m};\mathbb{C})\backslash \{0\}$ and $\Psi
_{b}\in \mathbb{C}\backslash \{0\}$. By combining the two equations of (\ref%
{eq super macarena}) we arrive at the assertion. \hfill $\Box $

We next analyze conditions for the existence of negative eigenvalues of $%
A(k) $ for $k\in \lbrack -\pi ,\pi )^{2}$. With this aim, we use the
Birman--Schwinger principle (Proposition \ref{lem-2.2}) to transform the
eigenvalue problem (\ref{eq super macarena}) into a non--linear equation for
$\lambda $ on negative reals. This permits us to study afterwards the
behavior of negative eigenvalues of $A(k)$ as functions of the couplings $%
\hat{v}$ and $U$.

\begin{proposition}[Negative eigenvalues of $A(k)$ -- II]
\label{effective BCS copy(1)}\mbox{ }\newline
Let $k\in \lbrack -\pi ,\pi )^{2}$ and $\lambda <0$. Then, $\lambda \in
\sigma (A(k))$ if and only if%
\begin{equation}
\left\vert \hat{v}\left( k\right) \right\vert ^{2}\mathcal{R}(k,U,\lambda
)+\lambda -\epsilon h_{b}(2-\cos (k))=0\ ,  \label{solution of balaaoao}
\end{equation}%
where%
\begin{equation}
\mathcal{R}(k,U,\lambda ):=\left\langle \mathfrak{d}(k)+\mathfrak{s}%
,(A_{1,1}(k)-\lambda )^{-1}(\mathfrak{d}(k)+\mathfrak{s)}\right\rangle \ .
\label{definition de R}
\end{equation}%
In this case, $\lambda $ is a non--degenerated discrete eigenvalue of $A(k)$.
\end{proposition}

\noindent \textit{Proof.} Fix $k\in \lbrack -\pi ,\pi )^{2}$. Since%
\begin{equation}
\left\Vert \mathfrak{d}(k)+\mathfrak{s}\right\Vert _{2}^{2}=\left\Vert
\mathfrak{d}(k)\right\Vert _{2}^{2}+\left\Vert \mathfrak{s}\right\Vert
_{2}^{2}=4\mathrm{e}^{-2\kappa }+1\ ,  \label{alegria alegria}
\end{equation}%
note from (\ref{A legallll2})--(\ref{A legallll3}) that%
\begin{equation*}
A_{1,2}(k)A_{2,1}(k)=\left( 1+4\mathrm{e}^{-2\kappa }\right) \left\vert \hat{%
v}\left( k\right) \right\vert ^{2}P_{\mathfrak{d}(k)}\ ,
\end{equation*}%
where $P_{\mathfrak{d}(k)}$ is the orthogonal projection with range (\ref%
{range}). Because $U,\epsilon ,h_{b}\geq 0$, recall that $A_{1,1}(k)\geq 0$
and $A_{2,2}(k)\geq 0$, see (\ref{A legallll1})--(\ref{A legall0}) and (\ref%
{A legallll4}). Hence, by applying Lemma \ref{effective BCS} and Proposition %
\ref{lem-2.2} (Birman--Schwinger principle) to
\begin{equation*}
H_{0}=A_{1,1}(k)\text{\qquad and\qquad }V=A_{1,2}(k)(A_{2,2}(k)-\lambda
)^{-1}A_{2,1}(k)\ ,
\end{equation*}%
$\lambda <0$ is an eigenvalue of $A(k)$ if and only if%
\begin{equation}
\left( 1+4\mathrm{e}^{-2\kappa }\right) \left\vert \hat{v}\left( k\right)
\right\vert ^{2}P_{\mathfrak{d}(k)}(A_{1,1}(k)-\lambda )^{-1}P_{\mathfrak{d}%
(k)}=(\epsilon h_{b}(2-\cos (k))-\lambda )P_{\mathfrak{d}(k)}\ .
\label{equation briman gshow}
\end{equation}%
Note that $P_{\mathfrak{d}(k)}$ is a rank one projector and hence,\ again by
Proposition \ref{lem-2.2}, $\lambda $ is a non--degenerated eigenvalue of $%
A(k)$. So, we deduce the assertion from (\ref{alegria alegria}), (\ref%
{equation briman gshow}) and the fact that $\lambda \in \sigma (A(k))$ with $%
\lambda <0$ implies $\lambda \in \sigma _{\mathrm{d}}(A(k))$, by (\ref{ess
spectrum}).\hfill $\Box $

We next study the behavior of the function $\mathcal{R}(k,U,\lambda )$ for
negative spectral parameters $\lambda <0$ at any $k\in \lbrack -\pi ,\pi
)^{2}$.

\begin{lemma}[Behavior of the function $\protect\lambda \mapsto \mathcal{R}%
(k,U,\protect\lambda )$]
\label{effective BCS copy(4)}\mbox{ }\newline
Let $k\in \lbrack -\pi ,\pi )^{2}$ and $U\geq 0$. Then
\begin{equation*}
\mathcal{R}(k,U,\lambda )=4\left\vert \lambda \right\vert ^{-1}\mathrm{e}%
^{-2\kappa }+(U+\left\vert \lambda \right\vert )^{-1}+\mathcal{S}%
(k,U,\lambda )
\end{equation*}%
is a strictly increasing function of $\lambda <0$ with%
\begin{equation*}
\left\vert \mathcal{S}(k,U,\lambda )\right\vert \leq 8\epsilon \lambda
^{-2}(1+4\mathrm{e}^{-2\kappa })\ .
\end{equation*}
\end{lemma}

\noindent \textit{Proof. }Fix in all the proof $k\in \lbrack -\pi ,\pi )^{2}$%
, $U\geq 0$ and $\lambda <0$. First, it is easy to check that the function $%
\lambda \mapsto \mathcal{R}(k,U,\lambda )$ is strictly increasing for
negative $\lambda <0$, by strict positivity of the operator $%
(A_{1,1}(k)-\lambda )^{-2}$ when $\epsilon ,U\geq 0$. Secondly, by using (%
\ref{A legallll1})--(\ref{A legall0}) and the second resolvent equation we
obtain%
\begin{equation}
(A_{1,1}(k)-\lambda )^{-1}=(UP_{0}-\lambda )^{-1}-(UP_{0}-\lambda
)^{-1}A_{1,1}^{(0)}(k)(A_{1,1}(k)-\lambda )^{-1}\ .
\label{baila tu cuerpo alegria macarena}
\end{equation}%
Clearly,
\begin{equation*}
\Vert (UP_{0}-\lambda )^{-1}\Vert _{\mathrm{op}},\Vert (A_{1,1}(k)-\lambda
)^{-1}\Vert _{\mathrm{op}}\leq |\lambda |^{-1}\quad \text{and}\quad \Vert
A_{1,1}^{(0)}(k)\Vert _{\mathrm{op}}\leq 8\epsilon \ .
\end{equation*}%
It follows that%
\begin{equation*}
\left\Vert (A_{1,1}(k)-\lambda )^{-1}-(UP_{0}-\lambda )^{-1}\right\Vert _{%
\mathrm{op}}\leq 8\epsilon \lambda ^{-2}\ .
\end{equation*}%
Therefore, from (\ref{definition de R}) and (\ref{alegria alegria}),
\begin{equation}
\left\vert \mathcal{R}(k,U,\lambda )-\left\langle \mathfrak{d}(k)+\mathfrak{s%
},(UP_{0}-\lambda )^{-1}(\mathfrak{d}(k)+\mathfrak{s})\right\rangle
\right\vert \leq 8\epsilon \lambda ^{-2}(1+4\mathrm{e}^{-2\kappa })\ .
\label{cool}
\end{equation}%
Meanwhile, recall that $P_{0}$ is the orthogonal projection with range (\ref%
{range0}) while $\left\langle \mathfrak{d}(k),\mathfrak{s}\right\rangle =0$.
Therefore,
\begin{eqnarray*}
&&\left\langle \mathfrak{d}(k)+\mathfrak{s},(UP_{0}-\lambda )^{-1}(\mathfrak{%
d}(k)+\mathfrak{s})\right\rangle \\
&=&\left\langle \mathfrak{d}(k),(UP_{0}-\lambda )^{-1}\mathfrak{d}%
(k)\right\rangle +\left\langle \mathfrak{s},(UP_{0}-\lambda )^{-1}\mathfrak{s%
}\right\rangle \\
&=&\left\vert \lambda \right\vert ^{-1}\left\langle \mathfrak{d}(k),%
\mathfrak{d}(k)\right\rangle +(U+\left\vert \lambda \right\vert
)^{-1}\left\langle \mathfrak{s},\mathfrak{s}\right\rangle \ ,
\end{eqnarray*}%
which, combined with (\ref{cool}), yields the assertion.\hfill $\Box $

From Proposition \ref{effective BCS copy(1)} and Lemma \ref{effective BCS
copy(4)} we deduce the possible existence of a unique negative eigenvalue:

\begin{theorem}[Estimates on the negative eigenvalue of $A\left( k\right) $]

\label{effective BCS copy(3)}\mbox{ }\newline
Let $k\in \lbrack -\pi ,\pi )^{2}$, $h_{b}\in \left[ 0,1\right] $, $U\geq 0$
and set%
\begin{equation*}
\epsilon _{0}:=\frac{\mathrm{e}^{-2\kappa }}{4(1+4\mathrm{e}^{-2\kappa })}%
\mathrm{e}^{-\kappa }\ ,\qquad \kappa >0\ .
\end{equation*}%
\newline
\emph{(i)} There is at most one negative eigenvalue $\mathrm{E}(k)<0$ of $%
A(k)$. If it exists, $\mathrm{E}(k)$ is non--degenerated. \newline
\emph{(ii)} If $0\leq \epsilon \leq \epsilon _{0}\left\vert \hat{v}\left(
k\right) \right\vert $ with $\left\vert \hat{v}\left( k\right) \right\vert
\neq 0$ then there is a negative eigenvalue $\mathrm{E}(k)\equiv \mathrm{E}%
(k,U,\epsilon )$ of $A(k)$ that satisfies%
\begin{equation*}
\mathrm{e}^{-\kappa }\left\vert \hat{v}\left( k\right) \right\vert
<\left\vert \mathrm{E}(k)\right\vert <\left\vert \hat{v}\left( k\right)
\right\vert \sqrt{1+5\mathrm{e}^{-2\kappa }}\ .
\end{equation*}
\end{theorem}

\noindent \textit{Proof.} (i) Use Proposition \ref{effective BCS copy(1)}
and the monotonicity of the map $\lambda \mapsto \mathcal{R}(k,U,\lambda )$
on $\mathbb{R}^{-}$ (Lemma \ref{effective BCS copy(4)}).

\noindent (ii) Fix $k\in \lbrack -\pi ,\pi )^{2}$, $h_{b}\in \left[ 0,1%
\right] $ and $U\geq 0$. Assume that $\left\vert \hat{v}\left( k\right)
\right\vert \neq 0$, let $x:=\mathrm{e}^{-\kappa }\left\vert \hat{v}\left(
k\right) \right\vert $ and take $\epsilon \geq 0$ such that%
\begin{equation}
0\leq \epsilon \leq \epsilon _{0}\left\vert \hat{v}\left( k\right)
\right\vert <\frac{x}{4}\ .  \label{blabla}
\end{equation}%
Then, by Lemma \ref{effective BCS copy(4)},
\begin{eqnarray*}
&&\left\vert \hat{v}\left( k\right) \right\vert ^{2}\mathcal{R}(k,U,-x)-2x \\
&=&\left\vert \hat{v}\left( k\right) \right\vert ^{2}\left( 4x^{-1}\mathrm{e}%
^{-2\kappa }+(U+x)^{-1}+\mathcal{S}(k,U,-x)\right) -2x \\
&\geq &\left\vert \hat{v}\left( k\right) \right\vert ^{2}4\left( x^{-1}%
\mathrm{e}^{-2\kappa }-2\epsilon x^{-2}(1+4\mathrm{e}^{-2\kappa })\right) -2x
\\
&\geq &\left\vert \hat{v}\left( k\right) \right\vert ^{2}2x^{-1}\mathrm{e}%
^{-2\kappa }-2x=0\ .
\end{eqnarray*}%
Recall now that $\mathcal{R}(k,U,\lambda )>0$ is a strictly positive and
increasing function of $\lambda <0$, by Lemma \ref{effective BCS copy(4)}.
Therefore, for any parameter $\epsilon $ satisfying (\ref{blabla}), there is
a solution $\mathrm{E}(k)$ of (\ref{solution of balaaoao}) that satisfies $%
\left\vert \mathrm{E}(k)\right\vert >x$. Note that one also uses here $%
h_{b}\in \left[ 0,1\right] $.

Now, take%
\begin{equation*}
y=\left\vert \hat{v}\left( k\right) \right\vert \sqrt{1+5\mathrm{e}%
^{-2\kappa }}\ ,
\end{equation*}%
where we recall that $\left\vert \hat{v}\left( k\right) \right\vert \neq 0$.
Then, using (\ref{blabla}) and Lemma \ref{effective BCS copy(4)}, we obtain
that
\begin{eqnarray*}
\left\vert \hat{v}\left( k\right) \right\vert ^{2}\mathcal{R}(k,U,-y)-y
&\leq &\left\vert \hat{v}\left( k\right) \right\vert ^{2}y^{-1}\left( 4%
\mathrm{e}^{-2\kappa }+1+8\epsilon y^{-1}(1+4\mathrm{e}^{-2\kappa })\right)
-y \\
&\leq &\frac{\left\vert \hat{v}\left( k\right) \right\vert \mathrm{e}%
^{-3\kappa }}{\left( 1+5\mathrm{e}^{-2\kappa }\right) }\left( 2-\sqrt{5+%
\mathrm{e}^{2\kappa }}\right) <0\ .
\end{eqnarray*}%
Then, by Lemma \ref{effective BCS copy(4)}, if $\epsilon $ satisfies (\ref%
{blabla}), there is a solution $\mathrm{E}(k)$ of (\ref{solution of balaaoao}%
) satisfying $\left\vert \mathrm{E}(k)\right\vert <y$. \hfill $\Box $

We now conclude this subsection by proving the existence of a bound fermion
pair whenever the bottom $E_{0}$ of the spectrum of $H^{(2,1)}$ is strictly
negative:

\begin{proposition}[Bound fermion pair formation at strictly negative energy
-- I]
\label{effective BCS copy(7)}Assume that $E_{0}<0$ and take any $(\mathfrak{c%
}_{0},\mathfrak{b}_{0})\in \mathfrak{H}_{\varepsilon }\backslash \{0\}$ with
$\varepsilon \in (0,1)$. Let
\begin{equation*}
(\mathfrak{c}_{t},\mathfrak{b}_{t}):=\mathrm{e}^{-itH^{(2,1)}}(\mathfrak{c}%
_{0},\mathfrak{b}_{0})\ ,\qquad t\in \mathbb{R}\ .
\end{equation*}%
\emph{(i)} Non--vanishing fermion component:
\begin{equation*}
\left\Vert \mathfrak{c}_{t}\right\Vert _{2}=\left\Vert \mathfrak{c}%
_{0}\right\Vert _{2}>0\ ,\qquad t\in \mathbb{R}\ .
\end{equation*}%
\emph{(ii)} Bound fermion pair formation:%
\begin{equation*}
\lim_{R\rightarrow \infty }\sum_{x_{\uparrow },x_{\downarrow }\in \mathbb{Z}%
^{2}:|x_{\uparrow }-x_{\downarrow }|\leq R}|\mathfrak{c}_{t}(x_{\uparrow
},x_{\downarrow })|^{2}=\left\Vert \mathfrak{c}_{0}\right\Vert _{2}^{2}\ ,
\end{equation*}%
\emph{uniformly} with respect to $t\in \mathbb{R}$.
\end{proposition}

\noindent \textit{Proof.} (i) For any $t\in \mathbb{R}$, let $(\mathfrak{%
\hat{c}}_{t},\mathfrak{\hat{b}}_{t}):=\mathfrak{U}^{\ast }(\mathfrak{c}_{t},%
\mathfrak{b}_{t})$, see (\ref{u stra}). If $E_{0}<0$ and $(\mathfrak{c}_{0},%
\mathfrak{b}_{0})\in \mathfrak{H}_{\varepsilon }$, then we infer from Lemma %
\ref{Carlos super main copy(5)},\ Equation (\ref{ess spectrum}) and Theorem %
\ref{effective BCS copy(3)} (i) together with Proposition \ref{proposition
Self--adjoint decomposable operators} (iii) that
\begin{equation}
(\mathfrak{c}_{t},\mathfrak{b}_{t})=\mathfrak{U}\left( \int_{[-\pi ,\pi
)^{2}}^{\oplus }\mathrm{e}^{-it\mathrm{E}(k)}(\mathfrak{\hat{c}}_{0}\left(
k\right) ,\mathfrak{\hat{b}}_{0}\left( k\right) )\text{ }\mathfrak{m}(%
\mathrm{d}^{2}k)\right)  \label{eq 1}
\end{equation}%
for any $t\in \mathbb{R}$, where, by Lemma \ref{effective BCS},
\begin{equation}
\mathfrak{\hat{c}}_{0}\left( k\right) =0\text{\qquad if and only if\qquad }%
\mathfrak{\hat{b}}_{0}\left( k\right) =0\ .  \label{eq 2}
\end{equation}%
This implies Assertion (i), because
\begin{equation}
\left\Vert \mathfrak{c}_{t}\right\Vert _{2}^{2}=\int_{[-\pi ,\pi
)^{2}}\left\Vert \mathfrak{\hat{c}}_{0}\left( k\right) \right\Vert _{2}^{2}%
\text{ }\mathfrak{m}(\mathrm{d}^{2}k)=\left\Vert \mathfrak{c}_{0}\right\Vert
_{2}^{2}\ .  \label{eq sup0}
\end{equation}

\noindent (ii) By (\ref{eq 1}), there is a family $\{P^{(R)}\}_{R\in \mathbb{%
R}^{+}}$ of orthogonal projectors acting on $L^{2}([-\pi ,\pi )^{2},%
\mathfrak{m};\mathbb{C})$, converging strongly to the identity as $%
R\rightarrow \infty $, and such that
\begin{equation}
\sum_{x_{\uparrow },x_{\downarrow }\in \mathbb{Z}^{2}:|x_{\uparrow
}-x_{\downarrow }|\leq R}|\mathfrak{c}_{t}(x_{\uparrow },x_{\downarrow
})|^{2}=\int_{[-\pi ,\pi )^{2}}\left\Vert P^{(R)}\mathfrak{\hat{c}}%
_{0}\left( k\right) \right\Vert _{2}^{2}\text{ }\mathfrak{m}(\mathrm{d}%
^{2}k)\ .  \label{eq sup}
\end{equation}%
By Lebesgue's dominated convergence theorem, Assertion (ii) then
follows.\hfill $\Box $

\subsection{Coefficients $\mathcal{R}(k,U,\protect\lambda )$ in terms of
Explicit Integrals}

To prove Theorem \ref{pairing mode copy(3)}, we need to express the positive
numbers $\mathcal{R}(k,U,\lambda )$ for $k\in \lbrack -\pi ,\pi )^{2}$, $%
U\geq 0$ and $\lambda <0$ in terms of the explicit quantities $R_{\mathfrak{s%
},\mathfrak{s}}^{(0)}$, $R_{\mathfrak{d},\mathfrak{d}}^{(0)}$, and $R_{%
\mathfrak{s},\mathfrak{d}}^{(0)}$ defined by (\ref{Rss})--(\ref{Rss2}). See
Proposition \ref{effective BCS copy(1)} for the definition and the use of $%
\mathcal{R}(k,U,\lambda )$. This is done in the following lemma:

\begin{lemma}[Explicit expression for $\mathcal{R}(k,U,\protect\lambda )$]
\label{effective BCS copy(2)bis}\mbox{ }\newline
Let $k\in \lbrack -\pi ,\pi )^{2}$, $U\geq 0$ and $\lambda <0$. Then
\begin{equation*}
\mathcal{R}(k,U,\lambda )=\frac{R_{\mathfrak{d},\mathfrak{d}}^{(0)}}{1+UR_{%
\mathfrak{s},\mathfrak{s}}^{(0)}}+U\frac{R_{\mathfrak{d},\mathfrak{d}%
}^{(0)}R_{\mathfrak{s},\mathfrak{s}}^{(0)}-(R_{\mathfrak{s},\mathfrak{d}%
}^{(0)})^{2}}{1+UR_{\mathfrak{s},\mathfrak{s}}^{(0)}}
\end{equation*}%
with%
\begin{equation}
\frac{R_{\mathfrak{d},\mathfrak{d}}^{(0)}}{1+UR_{\mathfrak{s},\mathfrak{s}%
}^{(0)}}>0\text{\qquad and\qquad }U\frac{R_{\mathfrak{d},\mathfrak{d}%
}^{(0)}R_{\mathfrak{s},\mathfrak{s}}^{(0)}-(R_{\mathfrak{s},\mathfrak{d}%
}^{(0)})^{2}}{1+UR_{\mathfrak{s},\mathfrak{s}}^{(0)}}\geq 0\ .
\label{postivity}
\end{equation}
\end{lemma}

\noindent \textit{Proof.} Fix $k\in \lbrack -\pi ,\pi )^{2}$, $U\geq 0$ and $%
\lambda <0$. From Definitions (\ref{Rss})--(\ref{Rss2}), (\ref{def d et s}),
and (\ref{A legallll1}), note that
\begin{eqnarray*}
R_{\mathfrak{s},\mathfrak{s}}^{(0)} &=&\left\langle \mathfrak{s}%
,(A_{1,1}^{(0)}(k)-\lambda )^{-1}\mathfrak{s}\right\rangle \ , \\
R_{\mathfrak{d},\mathfrak{d}}^{(0)} &=&\left\langle \mathfrak{d}(k)+%
\mathfrak{s},(A_{1,1}^{(0)}(k)-\lambda )^{-1}(\mathfrak{d}(k)+\mathfrak{s}%
)\right\rangle \ , \\
R_{\mathfrak{s},\mathfrak{d}}^{(0)} &=&\left\langle \mathfrak{s}%
,(A_{1,1}^{(0)}(k)-\lambda )^{-1}(\mathfrak{d}(k)+\mathfrak{s})\right\rangle
\ .
\end{eqnarray*}%
Define the real numbers%
\begin{eqnarray}
R_{\mathfrak{s},\mathfrak{s}} &:=&\left\langle \mathfrak{s},(A_{1,1}(k)-\lambda )^{-1}\mathfrak{s}\right\rangle \ ,  \notag \\
R_{\mathfrak{d},\mathfrak{d}} &:=&\left\langle \mathfrak{d}(k)+\mathfrak{s},(A_{1,1}(k)-\lambda )^{-1}(\mathfrak{d}(k)+\mathfrak{s})\right\rangle \ ,
\label{Rdd} \\
R_{\mathfrak{s},\mathfrak{d}} &:=&\left\langle \mathfrak{s},(A_{1,1}(k)-\lambda )^{-1}(\mathfrak{d}(k)+\mathfrak{s})\right\rangle \ .
\notag
\end{eqnarray}%
and observe that $R_{\mathfrak{d},\mathfrak{d}}$ is only another notation
for $\mathcal{R}(k,U,\lambda )$:
\begin{equation}
R_{\mathfrak{d},\mathfrak{d}}=\mathcal{R}(k,U,\lambda )\ .
\label{resolvent constant}
\end{equation}%
From the resolvent equation%
\begin{equation}
(A_{1,1}(k)-\lambda )^{-1}=(A_{1,1}^{(0)}(k)-\lambda
)^{-1}-U(A_{1,1}^{(0)}(k)-\lambda )^{-1}P_{0}(A_{1,1}(k)-\lambda )^{-1}
\label{resolvent equation}
\end{equation}%
(see (\ref{A legall0})), we arrive at the linear system%
\begin{eqnarray*}
\left(
\begin{array}{ll}
R_{\mathfrak{s},\mathfrak{s}} & R_{\mathfrak{s},\mathfrak{d}} \\
R_{\mathfrak{s},\mathfrak{d}} & R_{\mathfrak{d},\mathfrak{d}}%
\end{array}%
\right) &=&-U\left(
\begin{array}{ll}
R_{\mathfrak{s},\mathfrak{s}}^{(0)}R_{\mathfrak{s},\mathfrak{s}} & R_{%
\mathfrak{s},\mathfrak{s}}^{(0)}R_{\mathfrak{s},\mathfrak{d}} \\
R_{\mathfrak{s},\mathfrak{d}}^{(0)}R_{\mathfrak{s},\mathfrak{s}} & R_{%
\mathfrak{s},\mathfrak{d}}^{(0)}R_{\mathfrak{s},\mathfrak{d}}%
\end{array}%
\right) +\left(
\begin{array}{ll}
R_{\mathfrak{s},\mathfrak{s}}^{(0)} & R_{\mathfrak{s},\mathfrak{d}}^{(0)} \\
R_{\mathfrak{s},\mathfrak{d}}^{(0)} & R_{\mathfrak{d},\mathfrak{d}}^{(0)}%
\end{array}%
\right) \\
&=&-\left(
\begin{array}{ll}
UR_{\mathfrak{s},\mathfrak{s}}^{(0)} & 0 \\
UR_{\mathfrak{s},\mathfrak{d}}^{(0)} & 0%
\end{array}%
\right) \left(
\begin{array}{ll}
R_{\mathfrak{s},\mathfrak{s}} & R_{\mathfrak{s},\mathfrak{d}} \\
R_{\mathfrak{s},\mathfrak{d}} & R_{\mathfrak{d},\mathfrak{d}}%
\end{array}%
\right) +\left(
\begin{array}{ll}
R_{\mathfrak{s},\mathfrak{s}}^{(0)} & R_{\mathfrak{s},\mathfrak{d}}^{(0)} \\
R_{\mathfrak{s},\mathfrak{d}}^{(0)} & R_{\mathfrak{d},\mathfrak{d}}^{(0)}%
\end{array}%
\right) \ ,
\end{eqnarray*}%
which means%
\begin{equation*}
\left(
\begin{array}{ll}
UR_{\mathfrak{s},\mathfrak{s}}^{(0)}+1 & 0 \\
UR_{\mathfrak{s},\mathfrak{d}}^{(0)} & 1%
\end{array}%
\right) \left(
\begin{array}{ll}
R_{\mathfrak{s},\mathfrak{s}} & R_{\mathfrak{s},\mathfrak{d}} \\
R_{\mathfrak{s},\mathfrak{d}} & R_{\mathfrak{d},\mathfrak{d}}%
\end{array}%
\right) =\left(
\begin{array}{ll}
R_{\mathfrak{s},\mathfrak{s}}^{(0)} & R_{\mathfrak{s},\mathfrak{d}}^{(0)} \\
R_{\mathfrak{s},\mathfrak{d}}^{(0)} & R_{\mathfrak{d},\mathfrak{d}}^{(0)}%
\end{array}%
\right) \ .
\end{equation*}%
Note that, by positivity of the constants $R_{\mathfrak{s},\mathfrak{s}%
}^{(0)}\geq 0$ and $U\geq 0$, the matrix
\begin{equation*}
\left(
\begin{array}{ll}
UR_{\mathfrak{s},\mathfrak{s}}^{(0)}+1 & 0 \\
UR_{\mathfrak{s},\mathfrak{d}}^{(0)} & 1%
\end{array}%
\right)
\end{equation*}%
is invertible and we obtain%
\begin{equation*}
\left(
\begin{array}{ll}
R_{\mathfrak{s},\mathfrak{s}} & R_{\mathfrak{s},\mathfrak{d}} \\
R_{\mathfrak{s},\mathfrak{d}} & R_{\mathfrak{d},\mathfrak{d}}%
\end{array}%
\right) =\frac{1}{1+UR_{\mathfrak{s},\mathfrak{s}}^{(0)}}\left(
\begin{array}{ll}
1 & 0 \\
-UR_{\mathfrak{s},\mathfrak{d}}^{(0)} & UR_{\mathfrak{s},\mathfrak{s}%
}^{(0)}+1%
\end{array}%
\right) \left(
\begin{array}{ll}
R_{\mathfrak{s},\mathfrak{s}}^{(0)} & R_{\mathfrak{s},\mathfrak{d}}^{(0)} \\
R_{\mathfrak{s},\mathfrak{d}}^{(0)} & R_{\mathfrak{d},\mathfrak{d}}^{(0)}%
\end{array}%
\right) \ .
\end{equation*}%
In particular,%
\begin{equation*}
R_{\mathfrak{d},\mathfrak{d}}=\frac{R_{\mathfrak{d},\mathfrak{d}%
}^{(0)}(1+UR_{\mathfrak{s},\mathfrak{s}}^{(0)})-U(R_{\mathfrak{s},\mathfrak{d%
}}^{(0)})^{2}}{1+UR_{\mathfrak{s},\mathfrak{s}}^{(0)}}=\frac{R_{\mathfrak{d},%
\mathfrak{d}}^{(0)}}{1+UR_{\mathfrak{s},\mathfrak{s}}^{(0)}}+U\frac{R_{%
\mathfrak{d},\mathfrak{d}}^{(0)}R_{\mathfrak{s},\mathfrak{s}}^{(0)}-(R_{%
\mathfrak{s},\mathfrak{d}}^{(0)})^{2}}{1+UR_{\mathfrak{s},\mathfrak{s}}^{(0)}%
}\ ,
\end{equation*}%
which, combined with (\ref{resolvent constant}), implies the assertion. Note
that the second inequality of (\ref{postivity}) follows from the positivity
of the operator $(A_{1,1}^{(0)}(k)-\lambda )^{-1}$.\hfill $\Box $

The behavior of $\mathcal{R}(k,U,\lambda )$ at large Hubbard coupling $U\geq
0$ can now be deduced. This is useful for the proof of Theorem \ref{pairing
mode copy(3)} (ii).

\begin{lemma}[$\mathcal{R}(k,U,\protect\lambda )$ at large $U\geq 0$]
\label{effective BCS copy(2)+1}\mbox{ }\newline
For any $k\in \lbrack -\pi ,\pi )^{2}$, $U\geq 0$ and $\lambda <0$.
\begin{equation}
4\mathrm{e}^{-2\kappa }\left( \left\vert \lambda \right\vert +4\epsilon
\right) ^{-1}+(U+4\epsilon +\left\vert \lambda \right\vert )^{-1}\leq
\mathcal{R}(k,U,\lambda )\leq 4\mathrm{e}^{-2\kappa }\left\vert \lambda
\right\vert ^{-1}+(U+\left\vert \lambda \right\vert )^{-1}
\label{gshow facile}
\end{equation}%
and for any $U>0$,%
\begin{equation}
\left\vert \mathcal{R}(k,U,\lambda )-\frac{R_{\mathfrak{d},\mathfrak{d}%
}^{(0)}R_{\mathfrak{s},\mathfrak{s}}^{(0)}-(R_{\mathfrak{s},\mathfrak{d}%
}^{(0)})^{2}}{R_{\mathfrak{s},\mathfrak{s}}^{(0)}}\right\vert <(1+4\mathrm{e}%
^{-\kappa })^{2}U^{-1}\ .  \label{explicit bound}
\end{equation}
\end{lemma}

\noindent \textit{Proof.} The Inequalities (\ref{gshow facile}) are direct
consequences of the definition of $\mathcal{R}(k,U,\lambda )$ (see, e.g., (%
\ref{Rdd}) and (\ref{resolvent constant})) and the fact that $A^{-1}\leq
B^{-1}$ for any strictly positive operators $A,B>0$ with $B\leq A$.
Moreover, we infer from Lemma \ref{effective BCS copy(2)bis} that%
\begin{equation}
\mathcal{R}(k,U,\lambda )-\frac{R_{\mathfrak{d},\mathfrak{d}}^{(0)}R_{%
\mathfrak{s},\mathfrak{s}}^{(0)}-(R_{\mathfrak{s},\mathfrak{d}}^{(0)})^{2}}{%
R_{\mathfrak{s},\mathfrak{s}}^{(0)}}=\frac{(R_{\mathfrak{s},\mathfrak{d}%
}^{(0)})^{2}}{(1+UR_{\mathfrak{s},\mathfrak{s}}^{(0)})R_{\mathfrak{s},%
\mathfrak{s}}^{(0)}}\ .  \label{Gshow asymtotics1}
\end{equation}%
Note that $R_{\mathfrak{s},\mathfrak{s}}^{(0)}>0$, by strict positivity of $%
(A_{1,1}^{(0)}(k)-\lambda )^{-1}>0$. Since, by (\ref{Rss}) and (\ref{Rss2}),
\begin{equation*}
|R_{\mathfrak{s},\mathfrak{d}}^{(0)}|\leq \left( 1+4\mathrm{e}^{-\kappa
}\right) R_{\mathfrak{s},\mathfrak{s}}^{(0)}\ ,
\end{equation*}%
we thus deduce from (\ref{Gshow asymtotics1}) Inequality (\ref{explicit
bound}) for any $U>0$.\hfill $\Box $

\subsection{Pairing Mode of Fermions with Minimum Energy}

Recall that if at quasi--momentum $k\in \lbrack -\pi ,\pi )^{2}$,
\begin{equation*}
0\leq \epsilon \leq \epsilon _{0}\left\vert \hat{v}\left( k\right)
\right\vert =\frac{\mathrm{e}^{-2\kappa }}{4(1+4\mathrm{e}^{-2\kappa })}%
\mathrm{e}^{-\kappa }\left\vert \hat{v}\left( k\right) \right\vert \ ,
\end{equation*}%
then there is a unique negative eigenvalue $\mathrm{E}(k)\equiv \mathrm{E}%
(k,U,\epsilon )$ of $A(k)$ for any $h_{b}\in \left[ 0,1\right] $ and $U\geq
0 $. See Theorem \ref{effective BCS copy(3)}. In this section, we are
interested in the asympotics of $\mathrm{E}(k)$ in the limits of small
kinetic terms $\epsilon \rightarrow 0^{+}$ and large or small Hubbard
repulsions $U\rightarrow \infty ,0^{+}$. We start by general results which
hold for any $U\geq 0$, provided $\epsilon \leq \tilde{\epsilon}%
_{0}\left\vert \hat{v}\left( k\right) \right\vert $ with
\begin{equation}
\tilde{\epsilon}_{0}:=\frac{\mathrm{e}^{-2\kappa }}{4}\min \left\{ \frac{1}{6%
\sqrt{1+5\mathrm{e}^{-2\kappa }}},\frac{\mathrm{e}^{-\kappa }}{(1+4\mathrm{e}%
^{-2\kappa })}\right\} \leq \epsilon _{0}\ .  \label{proector estimate2}
\end{equation}

\begin{theorem}[Asymptotics of the negative eigenvalue of $A\left( k\right) $
-- I]
\label{pairing mode}\mbox{ }\newline
There is a constant $D_{\kappa }<\infty $\ depending only on $\kappa >0$
such that, for every $h_{b}\in \left[ 0,1\right] $, $U\geq 0$, $k\in \lbrack
-\pi ,\pi )^{2}$ and any parameter $\epsilon \geq 0$ satisfying $0\leq
\epsilon <\tilde{\epsilon}_{0}\left\vert \hat{v}\left( k\right) \right\vert $%
, one has: \newline
\emph{(i)} Negative (non--degenerated) eigenvalue of $A\left( k\right) $:%
\begin{equation*}
\left\vert \mathrm{E}(k)+\frac{\left\vert \hat{v}\left( k\right) \right\vert
}{2}\left( \frac{\left\vert \hat{v}\left( k\right) \right\vert }{U-\mathrm{E}%
(k)}+\sqrt{16\mathrm{e}^{-2\kappa }+\frac{\left\vert \hat{v}\left( k\right)
\right\vert ^{2}}{\left( U-\mathrm{E}(k)\right) ^{2}}}\right) \right\vert
\leq \epsilon D_{\kappa }\ .
\end{equation*}%
\emph{(ii)} Eigenvector: There is an eigenvector%
\begin{equation*}
(\Psi _{\uparrow \downarrow }(k),\Psi _{b}(k))\equiv (\Psi _{\uparrow
\downarrow }(k,U,\hat{v}),\Psi _{b}(k,U,\hat{v}))
\end{equation*}%
associated with $\mathrm{E}(k)$ such that%
\begin{equation*}
\left\Vert \Psi _{\uparrow \downarrow }(k)-\left( \frac{\mathrm{e}^{2\kappa }%
}{4}\left( \frac{\mathrm{E}(k)}{\hat{v}\left( k\right) }+\frac{\hat{v}\left(
k\right) }{U-\mathrm{E}(k)}\right) \mathfrak{d}(k)-\frac{\hat{v}\left(
k\right) }{U-\mathrm{E}(k)}\mathfrak{s}\right) \right\Vert _{2}\leq \epsilon
\left\vert \hat{v}\left( k\right) \right\vert ^{-1}D_{\kappa }
\end{equation*}%
and
\begin{equation}
\left\vert \Psi _{b}(k)-1\right\vert \leq \epsilon \left\vert \hat{v}\left(
k\right) \right\vert ^{-1}D_{\kappa }\ .  \label{bosonic part}
\end{equation}
\end{theorem}

\noindent \textit{Proof.} Similarly to $A(k)$, for any $U\geq 0$ and $k\in
\lbrack -\pi ,\pi )^{2}$, let%
\begin{equation*}
B(k):=\left(
\begin{array}{cc}
UP_{0} & A_{1,2}(k) \\
A_{2,1}(k) & 0%
\end{array}%
\right) \in \mathcal{B}(L^{2}([-\pi ,\pi )^{2},\mathfrak{m};\mathbb{C}%
)\times \mathbb{C}))\ ,
\end{equation*}%
where $P_{0}$ is the orthogonal projection with range (\ref{range0}).
Compare this with (\ref{A legallll1})--(\ref{A legallll5}) in the limit $%
\epsilon \rightarrow 0^{+}$ to see that $B(k)=A(k)|_{\epsilon =0}$.
Therefore, by Theorem \ref{effective BCS copy(3)}, if $\hat{v}\left(
k\right) \neq 0$ then there is a negative eigenvalue $\mathrm{F}(k)$ of $%
B(k) $ that satisfies%
\begin{equation}
\mathrm{e}^{-\kappa }\left\vert \hat{v}\left( k\right) \right\vert
<\left\vert \mathrm{F}(k)\right\vert <\left\vert \hat{v}\left( k\right)
\right\vert \sqrt{1+5\mathrm{e}^{-2\kappa }}\ .  \label{gshow a nouveau}
\end{equation}%
This eigenvalue is moreover non--degenerated and the unique strictly
negative eigenvalue of $B(k)$. In fact, it can be explicitly computed
together with its eigenvector.

Indeed, for any $k\in \lbrack -\pi ,\pi )^{2}$, let%
\begin{equation}
\Phi _{\uparrow \downarrow }(k):=\frac{\mathrm{e}^{2\kappa }}{4}\left( \frac{%
\mathrm{F}(k)}{\hat{v}\left( k\right) }+\frac{\hat{v}\left( k\right) }{U-%
\mathrm{F}(k)}\right) \mathfrak{d}(k)-\frac{\hat{v}\left( k\right) }{U-%
\mathrm{F}(k)}\mathfrak{s}\ .  \label{gshow a nouveau3}
\end{equation}%
Then, by explicit computations, observe that $(\Phi _{\uparrow \downarrow
}(k),1)$ is an eigenvector of $B(k)$ associated with the eigenvalue%
\begin{equation}
\mathrm{F}(k)=-\frac{\left\vert \hat{v}\left( k\right) \right\vert }{2}%
\left( \frac{\left\vert \hat{v}\left( k\right) \right\vert }{U-\mathrm{F}(k)}%
+\sqrt{16\mathrm{e}^{-2\kappa }+\left( \frac{\left\vert \hat{v}\left(
k\right) \right\vert }{U-\mathrm{F}(k)}\right) ^{2}}\right) <0\ .
\label{gshow a nouveau3+1}
\end{equation}%
Observe that a solution $\mathrm{F}(k)<0$ of (\ref{gshow a nouveau3+1})
always exists. Indeed, $\mathrm{F}(k)<0$ solves (\ref{gshow a nouveau3+1})
if and only if $\xi \left( \mathrm{F}(k)\right) =4\mathrm{e}^{-2\kappa }$,
where
\begin{equation}
\xi \left( x\right) :=\frac{x^{2}}{\left\vert \hat{v}\left( k\right)
\right\vert ^{2}}+\frac{x}{U-x}\ ,\qquad x<0\ .  \label{solve eq}
\end{equation}%
Note that $\xi \left( x\right) \rightarrow \infty $, as $x\rightarrow
-\infty $, and
\begin{equation*}
x_{0}=\frac{1}{2}\left( U-\sqrt{U^{2}+4\left\vert \hat{v}\left( k\right)
\right\vert ^{2}}\right)
\end{equation*}%
is the unique strictly negative solution of $\xi \left( x\right) =0$.
Therefore, by continuity of $\xi $ on $-\mathbb{R}_{0}^{+}$, there is a
solution $x_{1}<x_{0}$ of (\ref{solve eq}). Hence (\ref{gshow a nouveau3+1})
holds for some $\mathrm{F}(k)<0$.

By Theorem \ref{effective BCS copy(3)} (ii), if $\epsilon \leq \epsilon
_{0}\left\vert \hat{v}\left( k\right) \right\vert $ with $\hat{v}\left(
k\right) \neq 0$ then there is a strictly negative and non degenerated
eigenvalue $\mathrm{E}(k)$ of $A(k)$. This eigenvalue is close to $\mathrm{F}%
(k)$ for $\epsilon \ll 1$ because of the equality $B(k)=A(k)|_{\epsilon =0}$
and Kato's perturbation theory.

Indeed, let
\begin{equation}
\mathfrak{z}\left( k\right) :=\frac{1}{2}\left( \sqrt{1+5\mathrm{e}%
^{-2\kappa }}+\mathrm{e}^{-\kappa }\right) \ ,\qquad k\in \lbrack -\pi ,\pi
)^{2}\ ,  \label{gammak00}
\end{equation}%
and $\mathcal{C}$ be the contour defined by
\begin{equation}
\mathcal{C}\left( y\right) :=\left\vert \hat{v}\left( k\right) \right\vert
\left( -\mathfrak{z}\left( k\right) +\left( \mathfrak{z}\left( k\right) -%
\frac{\mathrm{e}^{-\kappa }}{2}\right) \mathrm{e}^{2\pi iy}\right) \in
\mathbb{C}\ ,\qquad y\in \lbrack 0,1]\ ,  \label{gammak0}
\end{equation}%
for any $k\in \lbrack -\pi ,\pi )^{2}$ with $\hat{v}\left( k\right) \neq 0$.
Then, we define the Riesz projections associated with $\mathrm{E}(k)$ and $%
\mathrm{F}(k)$ respectively by%
\begin{equation*}
\mathbf{P}^{(\mathrm{E}(k))}:=\frac{1}{2\pi i}\oint\limits_{\mathcal{C}%
}(\zeta -A(k))^{-1}\mathrm{d}\zeta \qquad \text{and}\qquad \mathbf{P}^{(%
\mathrm{F}(k))}:=\frac{1}{2\pi i}\oint\limits_{\mathcal{C}}(\zeta -B(k))^{-1}%
\mathrm{d}\zeta \ .
\end{equation*}%
Both operators are well--defined for any $h_{b}\in \left[ 0,1\right] $, $%
U\geq 0$, $k\in \lbrack -\pi ,\pi )^{2}$ with $\hat{v}\left( k\right) \neq 0$
and $\epsilon \leq \epsilon _{0}\left\vert \hat{v}\left( k\right)
\right\vert $ because of Theorem \ref{effective BCS copy(3)}, see also (\ref%
{gshow a nouveau}). Using the resolvent equation,%
\begin{equation*}
\mathbf{P}^{(\mathrm{E}(k))}-\mathbf{P}^{(\mathrm{F}(k))}=\frac{1}{2\pi i}%
\oint\limits_{\mathcal{C}}(\zeta -B(k))^{-1}\left( A(k)-B(k)\right) (\zeta
-A(k))^{-1}\mathrm{d}\zeta
\end{equation*}%
from which we deduce%
\begin{eqnarray}
\left\Vert \mathbf{P}^{(\mathrm{E}(k))}-\mathbf{P}^{(\mathrm{F}%
(k))}\right\Vert _{\mathrm{op}} &\leq &\frac{2\mathrm{e}^{2\kappa }\sqrt{1+5%
\mathrm{e}^{-2\kappa }}}{\left\vert \hat{v}\left( k\right) \right\vert }%
\left( \Vert A_{1,1}^{(0)}(k)\Vert _{\mathrm{op}}+\Vert A_{2,2}(k)\Vert _{%
\mathrm{op}}\right)  \notag \\
&\leq &24\mathrm{e}^{2\kappa }\sqrt{1+5\mathrm{e}^{-2\kappa }}\left\vert
\hat{v}\left( k\right) \right\vert ^{-1}\epsilon  \label{proector estimate}
\end{eqnarray}%
for any $h_{b}\in \left[ 0,1\right] $, $U\geq 0$, $k\in \lbrack -\pi ,\pi
)^{2}$ such that $\hat{v}\left( k\right) \neq 0$, and parameters $\epsilon
\leq \epsilon _{0}\left\vert \hat{v}\left( k\right) \right\vert $. See
Theorem \ref{effective BCS copy(3)} and Equations (\ref{A legallll1})--(\ref%
{A legallll5}), (\ref{ess spectrum}), (\ref{gshow a nouveau}) and (\ref%
{gammak00})--(\ref{gammak0}). Since
\begin{equation}
\mathbf{P}^{(\mathrm{F}(k))}(\Phi _{\uparrow \downarrow }(k),1)=(\Phi
_{\uparrow \downarrow }(k),1)\ ,
\end{equation}%
it follows that, for any $h_{b}\in \left[ 0,1\right] $, $U\geq 0$, $k\in
\lbrack -\pi ,\pi )^{2}$ such that $\hat{v}\left( k\right) \neq 0$, and
parameters $\epsilon <\tilde{\epsilon}_{0}\left\vert \hat{v}\left( k\right)
\right\vert $ (cf. (\ref{proector estimate2})), the vector
\begin{equation}
(\Psi _{\uparrow \downarrow }(k),\Psi _{b}(k)):=\mathbf{P}^{(\mathrm{E}%
(k))}(\Phi _{\uparrow \downarrow }(k),1)\neq 0  \label{proector estimate3}
\end{equation}%
is an eigenvector of $A(k)$ associated with the unique strictly negative
eigenvalue $\mathrm{E}(k)$ of $A(k)$ while
\begin{equation}
\left\vert \mathrm{E}(k)-\mathrm{F}(k)\right\vert \leq 12\epsilon \ .
\label{gshow a nouveau2}
\end{equation}%
By combining (\ref{proector estimate})--(\ref{gshow a nouveau2}) with
Theorem \ref{effective BCS copy(3)} (ii) and (\ref{gshow a nouveau3})--(\ref%
{gshow a nouveau3+1}) we arrive at the assertions from direct estimates.
Note only that $D_{\kappa }$ is a function of $\mathrm{e}^{2\kappa }$
exponentially growing to infinity when $\kappa \rightarrow \infty $. \hfill $%
\Box $

Clearly,%
\begin{equation*}
\left\Vert \Psi _{\uparrow \downarrow }(k)\right\Vert _{2}^{2}=1-\left\vert
\Psi _{b}(k)\right\vert ^{2}\leq 1
\end{equation*}%
is the probability of finding a pair of fermions, and not a boson, with
quasi--momentum $k\in \lbrack -\pi ,\pi )^{2}$. Similarly, $\left\vert \Psi
_{b}(k)\right\vert ^{2}$ is the probability of finding a boson with the same
quasi--momentum. The norm--one function $\left\Vert \Psi _{\uparrow
\downarrow }(k)\right\Vert _{2}^{-1}\Psi _{\uparrow \downarrow }(k)$
describes the orbital structure of the bound fermion pair. Theorem \ref%
{pairing mode} says that, for small parameters $\epsilon \leq \tilde{\epsilon%
}_{0}\left\vert \hat{v}\left( k\right) \right\vert $, the orbital structure
of the bound\ fermion pair has mainly $s$-- and $d$--wave components. This
fact holds true even in the limit $U\rightarrow 0^{+}$:

\begin{koro}[Asymptotics of the negative eigenvalue of $A\left( k\right) $
-- II]
\label{Theorem AC conductivity measure2 copy(2)}\mbox{
}\newline
There is a constant $D_{\kappa }<\infty $\ depending only on $\kappa >0$
such that, for every $h_{b}\in \left[ 0,1\right] $, $U\geq 0$, $k\in \lbrack
-\pi ,\pi )^{2}$ and any parameter $\epsilon \geq 0$ satisfying $0\leq
\epsilon <\tilde{\epsilon}_{0}\left\vert \hat{v}\left( k\right) \right\vert $%
, one has: \newline
\emph{(i)} Negative (non--degenerated) eigenvalue of $A\left( k\right) $:%
\begin{equation*}
\left\vert \mathrm{E}(k)+\left\vert \hat{v}\left( k\right) \right\vert \sqrt{%
1+4\mathrm{e}^{-2\kappa }}\right\vert \leq D_{\kappa }\left( \epsilon
+U\right) \ .
\end{equation*}%
\emph{(ii)} Eigenvector: There is an eigenvector $(\Psi _{\uparrow
\downarrow }(k),\Psi _{b}(k))$ associated with $\mathrm{E}(k)$ such that (%
\ref{bosonic part}) holds and%
\begin{equation*}
\left\Vert \Psi _{\uparrow \downarrow }(k)+\mathfrak{d}(k)+\mathfrak{s}%
\right\Vert _{2}\leq D_{\kappa }\left( \epsilon \left\vert \hat{v}\left(
k\right) \right\vert ^{-1}+U\right) \ .
\end{equation*}
\end{koro}

In contrast to $U\ll 1$, in the limit of large Hubbard couplings $U\gg 1$
the $s$--wave component of the orbital structure of the bound fermion pair
is suppressed by the Hubbard repulsion without changing (at leading order)
the binding energy of the particles.

\begin{koro}[Asymptotics of the negative eigenvalue of $A\left( k\right) $
-- III]
\label{Theorem AC conductivity measure2}\mbox{
}\newline
There is a constant $D_{\kappa }<\infty $\ depending only on $\kappa >0$
such that, for every $h_{b}\in \left[ 0,1\right] $, $U\geq 0$, $k\in \lbrack
-\pi ,\pi )^{2}$ and any parameter $\epsilon \geq 0$ satisfying $0\leq
\epsilon <\tilde{\epsilon}_{0}\left\vert \hat{v}\left( k\right) \right\vert $%
, one has: \newline
\emph{(i)} Negative (non--degenerated) eigenvalue of $A\left( k\right) $:%
\begin{equation*}
\left\vert \mathrm{E}(k)+2\mathrm{e}^{-\kappa }\left\vert \hat{v}\left(
k\right) \right\vert +\frac{\left\vert \hat{v}\left( k\right) \right\vert
^{2}}{2U}\right\vert \leq D_{\kappa }\left( \epsilon +U^{-2}\right) \ .
\end{equation*}%
\emph{(ii)} Eigenvector: There is an eigenvector $(\Psi _{\uparrow
\downarrow }(k),\Psi _{b}(k))$ associated with $\mathrm{E}(k)$ such that (%
\ref{bosonic part}) holds and%
\begin{equation*}
\left\Vert \Psi _{\uparrow \downarrow }(k)-\frac{\mathrm{e}^{\kappa }}{2}%
\mathrm{sgn}\left( \hat{v}\left( k\right) \right) \mathfrak{d}(k)-\frac{\hat{%
v}\left( k\right) }{U}\mathfrak{s}\right\Vert _{2}\leq D_{\kappa }\left(
\epsilon \left\vert \hat{v}\left( k\right) \right\vert ^{-1}+U^{-2}\right) \
.
\end{equation*}
\end{koro}

\noindent Note that Corollaries \ref{Theorem AC conductivity measure2
copy(2)}--\ref{Theorem AC conductivity measure2} and the operator
monotonicity of $A_{1,1}(k)$ with respect to $U$ imply that, for all $k\in
\lbrack -\pi ,\pi )^{2}$ with $\epsilon <\tilde{\epsilon}_{0}\left\vert \hat{%
v}\left( k\right) \right\vert $, and every $h_{b}\in \left[ 0,1\right] $ and
$U\geq 0$,
\begin{equation*}
2\mathrm{e}^{-\kappa }\left\vert \hat{v}\left( k\right) \right\vert +%
\mathcal{O}\left( \epsilon \right) \leq \left\vert \mathrm{E}(k)\right\vert
\leq \left\vert \hat{v}\left( k\right) \right\vert \sqrt{1+4\mathrm{e}%
^{-2\kappa }}+\mathcal{O}\left( \epsilon \right) \ .
\end{equation*}%
Compare with Theorem \ref{effective BCS copy(3)} (ii).

By definition of $\mathfrak{d}(k)$ (see (\ref{def d et s})), observe that%
\begin{equation*}
\left\Vert \frac{\mathrm{e}^{\kappa }}{2}\mathrm{sgn}\left( \hat{v}\left(
k\right) \right) \mathfrak{d}(k)\right\Vert _{2}=1
\end{equation*}%
for all $\kappa \in \mathbb{R}_{0}^{+}$. Hence, the fact that orbital of the
bound pair is of $d$--wave type only depends on $\left\vert \hat{v}\left(
k\right) \right\vert U^{-1}$ being small.

Recall that Proposition \ref{effective BCS copy(7)} shows the existence of a
bound\ fermion pair whenever the bottom $E_{0}$ of the spectrum of $%
H^{(2,1)} $ is strictly negative. In this case, for any $(\mathfrak{c}_{0},%
\mathfrak{b}_{0})\in \mathfrak{H}_{\varepsilon }\backslash \{0\}$ with $%
\varepsilon \in (0,1)$,
\begin{equation}
(\mathfrak{c}_{t},\mathfrak{b}_{t}):=\mathrm{e}^{-itH^{(2,1)}}(\mathfrak{c}%
_{0},\mathfrak{b}_{0})\ ,\qquad t\in \mathbb{R}\ ,  \label{solution}
\end{equation}%
has a non--vanishing fermion component, i.e., $\left\Vert \mathfrak{c}%
_{t}\right\Vert _{2}=\left\Vert \mathfrak{c}_{0}\right\Vert _{2}>0$. By
explicit computations, one checks that the $s$-- and $d$--wave components of
the orbital structure of the bound fermion pair both corresponds in the
lattice $\mathbb{Z}^{2}$ to wave functions with a fermion pair localized in
a ball of radius $1$. Therefore, by using Theorem \ref{pairing mode}, we can
improve\ Proposition \ref{effective BCS copy(7)} (ii):

\begin{koro}[Bound fermion pair formation at strictly negative energy - II]
\label{Theorem AC conductivity measure2 copy(1)}Assume $E_{0}<0$ and let $(%
\mathfrak{c}_{t},\mathfrak{b}_{t})$ be defined by (\ref{solution}) for any $%
t\in \mathbb{R}$ and $(\mathfrak{c}_{0},\mathfrak{b}_{0})\in \mathfrak{H}%
_{\varepsilon }\backslash \{0\}$ with $\varepsilon \in (0,1)$. Then, \emph{%
uniformly} with respect to $t\in \mathbb{R}$ and $(\mathfrak{c}_{0},%
\mathfrak{b}_{0})\in \mathfrak{H}_{\varepsilon }\backslash \{0\}$,%
\begin{equation*}
\lim_{\epsilon \rightarrow 0^{+}}\left\{ \left\Vert \mathfrak{c}%
_{0}\right\Vert _{2}^{-2}\sum_{x_{\uparrow },x_{\downarrow }\in \mathbb{Z}%
^{2}:|x_{\uparrow }-x_{\downarrow }|\leq 1}|\mathfrak{c}_{t}(x_{\uparrow
},x_{\downarrow })|^{2}\right\} =1\ .
\end{equation*}
\end{koro}

\noindent \textit{Proof.} From (\ref{eq sup}),
\begin{equation}
\sum_{x_{\uparrow },x_{\downarrow }\in \mathbb{Z}^{2}:|x_{\uparrow
}-x_{\downarrow }|\leq 1}|\mathfrak{c}_{t}(x_{\uparrow },x_{\downarrow
})|^{2}=\int_{[-\pi ,\pi )^{2}}\left\Vert P^{(1)}\mathfrak{\hat{c}}%
_{0}\left( k\right) \right\Vert _{2}^{2}\text{ }\mathfrak{m}(\mathrm{d}%
^{2}k)\ .  \label{eq sup2}
\end{equation}%
Theorems \ref{effective BCS copy(3)} (ii) and \ref{pairing mode} (ii) imply
the existence of $D<\infty $ such that, for all $k\in \lbrack -\pi ,\pi
)^{2} $ and $(\mathfrak{c}_{0},\mathfrak{b}_{0})\in \mathfrak{H}%
_{\varepsilon }\backslash \{0\}$,%
\begin{equation*}
\left\Vert \left( \mathbf{1}_{L^{2}([-\pi ,\pi )^{2},\mathfrak{m};\mathbb{C}%
)}-P^{(1)}\right) \mathfrak{\hat{c}}_{0}\left( k\right) \right\Vert _{2}\leq
\epsilon D\left\Vert \mathfrak{\hat{c}}_{0}\left( k\right) \right\Vert _{2}\
.
\end{equation*}%
By (\ref{eq sup0}) and (\ref{eq sup2}) together with Proposition \ref%
{effective BCS copy(7)} (i), the assertion then follows. \hfill $\Box $

The existence of the negative eigenvalue $\mathrm{E}(k)$ of $A\left(
k\right) $ is not clear in general. Therefore, we define the function $%
\mathrm{E}^{\mathrm{ext}}(k)$ for all $k\in \lbrack -\pi ,\pi )^{2}$ by $%
\mathrm{E}^{\mathrm{ext}}(k):=\mathrm{E}(k)$ if there is a negative
eigenvalue of $A\left( k\right) $\ and $\mathrm{E}^{\mathrm{ext}}(k):=0$
otherwise. To simplify notation, we set $\mathrm{E}^{\mathrm{ext}}(k)\equiv
\mathrm{E}(k)$. [By Kato's perturbation theory for the discrete spectrum of
closed operators together with Equation (\ref{ess spectrum}) and\ the
continuity of $\hat{v}$, the map $k\mapsto \mathrm{E}(k)$ from $[-\pi ,\pi
)^{2}$ to $\mathbb{R}_{0}^{-}$ is continuous. This information is not
important in the sequel.]

Recall (\ref{Kv}), that is,
\begin{equation*}
\mathfrak{K}_{v}:=\left\{ k\in \lbrack -\pi ,\pi )^{2}:|\hat{v}(k)|=\Vert
\hat{v}\Vert _{\infty }\right\} \ .
\end{equation*}%
This set can be seen as being the set of quasi--momenta of minimal energy,
up to some small errors when $\epsilon ^{-1},U\rightarrow \infty $:

\begin{lemma}[Quasi--momenta of minimal energy at large $\protect\epsilon %
^{-1},U$]
\label{effective BCS copy(2)}\mbox{ }\newline
Assume that $|\mathfrak{K}_{v}|<\infty $. For any $\eta >0$, there are $%
\varepsilon >0$ and $D<\infty $ such that, for all $\epsilon ^{-1},U\geq D$
and all $k\in \left[ -\pi ,\pi \right) ^{2}\backslash \left\{ \mathfrak{K}%
_{v}+B\left( 0,\eta \right) \right\} $,
\begin{equation*}
\mathrm{E}\left( k\right) \geq \left( 1-\varepsilon \right) \inf \mathrm{E}%
\left( \left[ -\pi ,\pi \right) ^{2}\right) \ .
\end{equation*}
\end{lemma}

\noindent \textit{Proof.} Assume without loss of generality that $\Vert \hat{%
v}\Vert _{\infty }>0$. For any $\eta >0$, there is $\varepsilon >0$ such
that, for all $k\in \left[ -\pi ,\pi \right) ^{2}\backslash \left\{
\mathfrak{K}_{v}+B\left( 0,\eta \right) \right\} $,
\begin{equation}
\left\vert \hat{v}\left( k\right) \right\vert \leq \left( 1-\varepsilon
\right) \max_{k\in \left[ -\pi ,\pi \right) ^{2}}\left\vert \hat{v}\left(
k\right) \right\vert \ ,  \label{holdsbis}
\end{equation}%
by continuity of $\hat{v}$. Indeed, assume the existence of $\eta >0$ such
that, for all $\varepsilon >0$, there would exist $k_{\varepsilon }\in \left[
-\pi ,\pi \right) ^{2}\backslash \left\{ \mathfrak{K}_{v}+B\left( 0,\eta
\right) \right\} $ so that
\begin{equation*}
\left\vert \hat{v}\left( k_{\varepsilon }\right) \right\vert >\left(
1-\varepsilon \right) \max_{k\in \left[ -\pi ,\pi \right) ^{2}}\left\vert
\hat{v}\left( k\right) \right\vert \ .
\end{equation*}%
By compacticity of $[-\pi ,\pi ]^{2}$ and continuity of $\hat{v}$, there is $%
k_{0}\in \left[ -\pi ,\pi \right) ^{2}\backslash \left\{ \mathfrak{K}%
_{v}+B\left( 0,\eta \right) \right\} $ with
\begin{equation*}
\left\vert \hat{v}\left( k_{0}\right) \right\vert =\max_{k\in \left[ -\pi
,\pi \right) ^{2}}\left\vert \hat{v}\left( k\right) \right\vert \ .
\end{equation*}%
Therefore, for any $\eta >0$, there is $\varepsilon >0$ such that, for all $%
k\in \left[ -\pi ,\pi \right) ^{2}\backslash \left\{ \mathfrak{K}%
_{v}+B\left( 0,\eta \right) \right\} $, (\ref{holdsbis}) holds true, which,
combined with Corollary \ref{Theorem AC conductivity measure2} (i), yields
the assertion. \hfill $\Box $

For any $k\in \left[ -\pi ,\pi \right) ^{2}$, let $\tilde{P}_{\mathfrak{d}%
(k)}$ be the orthogonal projection acting on the Hilbert space $L^{2}([-\pi
,\pi )^{2},\mathfrak{m};\mathbb{C})\times \mathbb{C}\ $with%
\begin{equation*}
\mathrm{Ran}(\tilde{P}_{\mathfrak{d}(k)})=\mathbb{C}\left( \frac{\mathrm{e}%
^{\kappa }}{2}\mathfrak{d}(k),\mathrm{sgn}\left( \hat{v}\left( k\right)
\right) \right) \ .
\end{equation*}%
Recall that $\mathrm{e}^{\kappa }=2\Vert \mathfrak{d}(k)\Vert _{2}^{-1}$.
Then, for all $\varepsilon >0$, define the projections
\begin{equation*}
\mathcal{P}_{\varepsilon }:=\int_{[-\pi ,\pi )^{2}}^{\oplus }\mathbf{1}%
_{[E_{0},E_{0}\left( 1-\varepsilon \right) ]}(\mathrm{E}\left( k\right) )\
\tilde{P}_{\mathfrak{d}(k)}\ \mathfrak{m}(\mathrm{d}^{2}k)
\end{equation*}%
and, for any $k\in \left[ -\pi ,\pi \right) ^{2}$ and $\eta ,\varepsilon >0$%
,
\begin{equation}
\mathcal{\tilde{P}}_{\varepsilon ,\eta }(k):=\int_{[-\pi ,\pi )^{2}}^{\oplus
}\chi _{\varepsilon ,\eta }^{(k)}(q)\ \tilde{P}_{\mathfrak{d}(k)}\ \mathfrak{%
m}(\mathrm{d}^{2}q)\ .  \label{projectino}
\end{equation}%
where, for any $q\in \left[ -\pi ,\pi \right) ^{2}$,%
\begin{equation}
\chi _{\varepsilon ,\eta }^{(k)}(q):=\mathbf{1}_{[E_{0},E_{0}\left(
1-\varepsilon \right) ]}(\mathrm{E}\left( q\right) )\mathbf{1}_{k+B\left(
0,\eta \right) }(q)\ .  \label{projectino2}
\end{equation}%
These operators are used to approximate now the spectral projection
\begin{equation*}
\mathbf{1}_{[E_{0},E_{0}\left( 1-\varepsilon \right) ]}(H^{(2,1)})
\end{equation*}%
of the Hamiltonian $H^{(2,1)}$ on the bottom $[E_{0},E_{0}\left(
1-\varepsilon \right) ]$ of its spectrum for any parameter $\varepsilon \in
\left( 0,1\right) $.

\begin{proposition}[Approximating projectors]
\label{projector approx}\mbox{ }\newline
Let $\varepsilon \in \left( 0,1\right) $ and assume that $E_{0}<0$. For any $%
\eta >0$, there is a constant $D<\infty $ such that, for all $\epsilon
^{-1},U\geq D$,
\begin{equation*}
\left\Vert \mathbf{1}_{[E_{0},E_{0}\left( 1-\varepsilon \right)
]}(H^{(2,1)})-\mathcal{P}_{\varepsilon }\right\Vert _{\mathrm{op}}\leq \eta
\ .
\end{equation*}%
Moreover, if $|\mathfrak{K}_{v}|<\infty $ and $\varepsilon \ll 1$ is
sufficiently small, then
\begin{equation*}
\left\Vert \mathbf{1}_{[E_{0},E_{0}\left( 1-\varepsilon \right)
]}(H^{(2,1)})-\sum_{k\in \mathfrak{K}_{v}}\mathcal{\tilde{P}}_{\varepsilon
,\eta }(k)\right\Vert _{\mathrm{op}}\leq \eta \ .
\end{equation*}
\end{proposition}

\noindent \textit{Proof.} By Lemma \ref{Carlos super main copy(4)} and
Proposition \ref{proposition Self--adjoint decomposable operators} (iii),
\begin{equation}
\mathbf{1}_{[E_{0},E_{0}\left( 1-\varepsilon \right)
]}(H^{(2,1)})=\int_{[-\pi ,\pi )^{2}}^{\oplus }\mathbf{1}_{[E_{0},E_{0}%
\left( 1-\varepsilon \right) ]}(A\left( k\right) )\ \mathfrak{m}(\mathrm{d}%
^{2}k)\ .  \label{**}
\end{equation}%
and the assertion follows by using Theorem \ref{effective BCS copy(3)},
Corollary \ref{Theorem AC conductivity measure2} (ii) and Lemma \ref%
{effective BCS copy(2)}. Note that $E_{0}<0$ yields $\Vert \hat{v}\Vert
_{\infty }>0$, by Lemma \ref{effective BCS}.\hfill $\Box $

Recall that the function $\mathbf{s}_{k}:\mathbb{Z}^{2}\rightarrow \mathbb{C}
$ is defined, for any $k\in \lbrack -\pi ,\pi )^{2}$, by (\ref{Sk}), that is,%
\begin{equation*}
\mathbf{s}_{k}\left( y\right) :=\frac{1}{2}%
\Big (%
\mathrm{e}^{ik\cdot (0,1)}\delta _{y,(0,1)}+\mathrm{e}^{ik\cdot
(0,-1)}\delta _{y,(0,-1)}+\mathrm{e}^{ik\cdot (1,0)}\delta _{y,(1,0)}+%
\mathrm{e}^{ik\cdot (-1,0)}\delta _{y,(-1,0)}%
\Big )%
\end{equation*}%
for all $y\in \mathbb{Z}^{2}$. For any $\eta ,\varepsilon >0$, $k\in \left[
-\pi ,\pi \right) ^{2}$ and $y\in \mathbb{Z}^{2}$, let%
\begin{eqnarray*}
g_{\varepsilon ,\eta }^{(k)}(y) &:=&\frac{1}{2}\int_{[-\pi ,\pi )^{2}}\mathfrak{m}(\mathrm{d}^{2}q)\ \chi _{\varepsilon ,\eta }^{(k)}(q)\mathrm{e}^{iq\cdot y}\int_{[-\pi ,\pi )^{2}}\mathfrak{m}(\mathrm{d}^{2}p)\cos (p-k)\left[ \mathfrak{\hat{c}}(q)\right] (p) \\
&&+\frac{\mathrm{sgn}\left( \hat{v}\left( k\right) \right) }{2}\int_{[-\pi
,\pi )^{2}}\mathfrak{m}(\mathrm{d}^{2}q)\ \chi _{\varepsilon ,\eta
}^{(k)}(q)\ \mathrm{e}^{iq\cdot y}\ \mathfrak{\hat{b}}\left( q\right)
\end{eqnarray*}%
with $\chi _{\varepsilon ,\eta }^{(k)}$ defined by (\ref{projectino2}).
These functions are important because they are directly related to the
projections (\ref{projectino}):

\begin{lemma}[Range of projections $\mathcal{\tilde{P}}_{\protect\varepsilon %
,\protect\eta }$]
\label{effective BCS copy(5)}\mbox{ }\newline
For any $(\mathfrak{c},\mathfrak{b})\in \mathcal{H}_{\uparrow ,\downarrow
}^{(2,1)}$, $\eta ,\varepsilon >0$, $k\in \left[ -\pi ,\pi \right) ^{2}$,
\begin{equation*}
\left[ \mathfrak{U}^{\ast }\mathcal{\tilde{P}}_{\varepsilon ,\eta }(k)%
\mathfrak{U}\right] (\mathfrak{c},\mathfrak{b})=(\mathfrak{c}^{\prime },%
\mathfrak{b}^{\prime })\ ,
\end{equation*}%
where, for any $x_{\uparrow },x_{\downarrow }\in \mathbb{Z}^{2}$,
\begin{equation*}
\mathfrak{c}^{\prime }(x_{\uparrow },x_{\downarrow })=\mathbf{s}%
_{k}(x_{\uparrow }-x_{\downarrow })g_{\varepsilon ,\eta }^{(k)}(x_{\uparrow
})
\end{equation*}%
and, for any $x_{b}\in \mathbb{Z}^{2}$,
\begin{eqnarray*}
\mathfrak{b}^{\prime }(x_{b}) &=&\frac{1}{2}\int_{[-\pi ,\pi )^{2}}\mathfrak{%
m}(\mathrm{d}^{2}q)\ \mathrm{e}^{iq\cdot x_{b}}\ \chi _{\varepsilon ,\eta
}^{(k)}(q)\ \mathfrak{\hat{b}}(q)+\frac{\mathrm{sgn}\left( \hat{v}\left(
k\right) \right) }{2}\int_{[-\pi ,\pi )^{2}}\mathfrak{m}(\mathrm{d}^{2}q) \\
&&\times \mathrm{e}^{iq\cdot x_{b}}\chi _{\varepsilon ,\eta
}^{(k)}(q)\int_{[-\pi ,\pi )^{2}}\mathfrak{m}(\mathrm{d}^{2}k_{\uparrow
\downarrow })\ \cos (k_{\uparrow \downarrow }-k)\left[ \mathfrak{\hat{c}}(q)%
\right] (k_{\uparrow \downarrow })\ .
\end{eqnarray*}
\end{lemma}

\noindent \textit{Proof.} Fix all the parameters of the lemma. We then
compute from (\ref{projectino})--(\ref{projectino2}) that%
\begin{multline*}
\mathfrak{c}^{\prime }(x_{\uparrow },x_{\downarrow })=\frac{1}{2}\left(
\int_{[-\pi ,\pi )^{2}}\mathrm{e}^{ik_{\uparrow \downarrow }\cdot
(x_{\downarrow }-x_{\uparrow })}\cos (k_{\uparrow \downarrow }-k)\ \mathfrak{%
m}(\mathrm{d}^{2}k_{\uparrow \downarrow })\right) \\
\times \int_{\lbrack -\pi ,\pi )^{2}}\mathfrak{m}(\mathrm{d}^{2}q)\ \chi
_{\varepsilon ,\eta }^{(k)}(q)\ \mathrm{e}^{iq\cdot x_{\uparrow }} \\
\left( \mathrm{sgn}\left( \hat{v}\left( k\right) \right) \ \mathfrak{\hat{b}}%
\left( q\right) +\int_{[-\pi ,\pi )^{2}}\mathfrak{m}(\mathrm{d}^{2}p)\cos
(p-k)\left[ \mathfrak{\hat{c}}(q)\right] (p)\right) \ ,
\end{multline*}%
while, for all $y\in \mathbb{Z}^{2}$,
\begin{equation}
\int_{\lbrack -\pi ,\pi )^{2}}\mathrm{e}^{ik_{\uparrow \downarrow }\cdot
y}\cos (k_{\uparrow \downarrow }-k)\ \mathfrak{m}(\mathrm{d}^{2}k_{\uparrow
\downarrow })=\mathbf{s}_{k}\left( y\right) \ ,
\label{fourier transform d wave}
\end{equation}%
by using (\ref{cosinus}). A similar computation can be done for $\mathfrak{b}%
^{\prime }$. We omit the details. \hfill $\Box $

For any $k\in \left[ -\pi ,\pi \right) ^{2}$, let the orthogonal projection $%
P_{\mathfrak{h}_{2,-}^{(0)}}$ acting on $\mathcal{H}_{\uparrow ,\downarrow
}^{(2,1)}\ $with range $\mathfrak{h}_{2,-}^{(0)}$. Recall that $\mathfrak{h}%
_{2,-}^{(0)}$ is the subspace of one zero--spin electron pair, which is
canonically isomorphic to the spaces (\ref{cooper pair space}). Then, we can
see a 50\% depletion of either the fermion pair density or the boson density
for large $\epsilon ^{-1},U$ (cf. (\ref{depletion})):

\begin{lemma}[Bosonic depletion at large $\protect\epsilon ^{-1},U$]
\label{effective BCS copy(6)}\mbox{ }\newline
Let $\varepsilon \in \left( 0,1\right) $ and assume that $E_{0}<0$. Then,
uniformly for all normalized vectors $\psi \in \mathcal{H}_{\uparrow
,\downarrow }^{(2,1)}$,
\begin{multline*}
\lim_{\epsilon ^{-1},U\rightarrow \infty }\left\vert \Vert P_{\mathfrak{h}%
_{2,-}^{(0)}}\mathbf{1}_{[E_{0},E_{0}\left( 1-\varepsilon \right)
]}(H^{(2,1)})\psi \Vert _{\mathcal{H}_{\uparrow ,\downarrow }^{(2,1)}}-\frac{%
1}{\sqrt{2}}\Vert \mathbf{1}_{[E_{0},E_{0}\left( 1-\varepsilon \right)
]}(H^{(2,1)})\psi \Vert _{\mathcal{H}_{\uparrow ,\downarrow
}^{(2,1)}}\right\vert \\
=0\ .
\end{multline*}
\end{lemma}

\noindent \textit{Proof.} It is direct consequence of Corollary \ref{Theorem
AC conductivity measure2} (ii) and Equation (\ref{**}).\hfill $\Box $

\subsection{Effective Fermi Model}

Similar to the three--body case studied above, the fermionic effective
Hamiltonian
\begin{equation*}
\tilde{H}^{(2,1)}:=\overline{\mathbf{\tilde{H}}|_{\mathcal{H}_{\uparrow
,\downarrow }^{(2,1)}}}\ ,
\end{equation*}%
which is defined by (\ref{uncoupled model}), is decomposable:%
\begin{equation}
\mathfrak{U}^{\ast }\tilde{H}^{(2,1)}\mathfrak{U}=\int_{[-\pi ,\pi
)^{2}}^{\oplus }\tilde{A}_{1,1}(k)\oplus \tilde{A}_{2,2}(k)\text{ }\mathfrak{%
m}(\mathrm{d}^{2}k)\ ,  \label{fiber decomposition}
\end{equation}%
where, for any $k\in \lbrack -\pi ,\pi )^{2}$,%
\begin{equation}
\tilde{A}_{1,1}(k):=A_{1,1}(k)-\left( 1+4\mathrm{e}^{-2\kappa }\right) \hat{w%
}_{f}\left( k\right) P_{\mathfrak{d}(k)}\ ,\quad \tilde{A}_{2,2}(k):=-\hat{w}%
_{b}(k)\ ,  \label{fiber decomposition2}
\end{equation}%
with $\hat{w}_{b}$ and $\hat{w}_{f}$ being the Fourier transforms of $w_{b}$
(\ref{v effecboson}) and $w_{f}$ (\ref{v effec}), respectively. See also (%
\ref{A legall0}) for the definition of $A_{1,1}(k)$ and recall that $P_{%
\mathfrak{d}(k)}$ is the orthogonal projection with range (\ref{range}). The
fiber decomposition of $\mathfrak{U}^{\ast }\tilde{H}^{(2,1)}\mathfrak{U}$
is obtained by direct computations and we omit the details.

The bosonic and fermionic subspaces are clearly invariant under the action
of $\tilde{H}^{(2,1)}$. Eigenvalues and eigenvectors of $\tilde{A}_{2,2}(k)$%
, $k\in \lbrack -\pi ,\pi )^{2}$, do not need to be discussed as these
operators act on one--dimensional spaces. Thus, we focus on the fermionic
subspace. Here, by (\ref{v effec}), $\hat{w}_{f}\left( k\right) \geq 0$ and
we can use the Birman--Schwinger principle (Proposition \ref{lem-2.2}) once
again with%
\begin{equation*}
H_{0}=A_{1,1}(k)\text{\qquad and\qquad }V=\left( 1+4\mathrm{e}^{-2\kappa
}\right) \hat{w}_{f}\left( k\right) P_{\mathfrak{d}(k)}\ ,
\end{equation*}%
to study the negative eigenvalues of $\tilde{A}_{1,1}(k)$: Let $k\in \lbrack
-\pi ,\pi )^{2}$. Then, $\lambda <0$ is an eigenvalue of $\tilde{A}_{1,1}(k)$
if and only if
\begin{equation}
\hat{w}_{f}\left( k\right) \mathcal{R}(k,U,\lambda )=1  \label{estimate}
\end{equation}%
with $\mathcal{R}(k,U,\lambda )$ defined by (\ref{definition de R}).
Moreover, this eigenvalue is non--degenerated and unique, by Lemma \ref%
{effective BCS copy(4)}. Note that, by compacticity of $P_{\mathfrak{d}(k)}$
and $\tilde{A}_{2,2}(k)$ as well as the positivity of $A_{1,1}(k)$, any
strictly negative eigenvalue of $\tilde{A}_{2,2}(k)$ is discrete. Comparing
the last equation with Proposition \ref{effective BCS copy(1)} and Corollary %
\ref{Theorem AC conductivity measure2}, $w_{f}$ (\ref{v effec}) is chosen
such that the negative eigenvalues $\mathrm{\tilde{E}}(k)$ and $\mathrm{E}%
(k) $ of $\tilde{A}_{1,1}(k)$ and $A(k)$, respectively, coincide in the
limit $\epsilon \rightarrow 0^{+}$\ and $U\rightarrow \infty $. Indeed, we
tune the parameter $\gamma _{f}>0$ in (\ref{v effec}) in order to maximize
the rate of convergence of
\begin{equation*}
|\mathrm{\tilde{E}}(k)-\mathrm{E}(k)|\rightarrow 0\text{ },
\end{equation*}%
as $\epsilon \rightarrow 0^{+}$, $U\rightarrow \infty $, and we obtain the
following result:

\begin{proposition}[Asymptotics of the negative eigenvalue of $\tilde{A}%
_{1,1}(k)$]
\label{pairing mode copy(1)}\mbox{ }\newline
Let $\gamma _{f}=\mathrm{e}^{\kappa }/2$. Then, there is a constant $%
D_{\kappa }<\infty $\ depending only on $\kappa >0$\ such that, for all $%
k\in \lbrack -\pi ,\pi )^{2}$, $\epsilon \geq 0$ satisfying $2\epsilon \leq
\tilde{\epsilon}_{0}\left\vert \hat{v}\left( k\right) \right\vert $, and
every $U\geq 0$, one has: \newline
\emph{(i)} There is a unique negative eigenvalue $\mathrm{\tilde{E}}(k)$ of $%
\tilde{A}_{1,1}(k)$. Moreover, it is non--degenerated and satisfies
\begin{equation*}
\left\vert \mathrm{\tilde{E}}(k)+2\mathrm{e}^{-\kappa }\left\vert \hat{v}%
\left( k\right) \right\vert +\frac{\left\vert \hat{v}\left( k\right)
\right\vert ^{2}}{2U}\right\vert \leq D_{\kappa }\left( \epsilon
+U^{-2}\right) \ .
\end{equation*}%
\emph{(ii)} There is an eigenvector $\tilde{\Psi}_{\uparrow \downarrow }(k)$
associated with $\mathrm{\tilde{E}}(k)$ satisfying
\begin{equation*}
\left\Vert \tilde{\Psi}_{\uparrow \downarrow }(k)-\frac{\mathrm{e}^{\kappa }%
}{2}\mathrm{sgn}\left( \hat{v}\left( k\right) \right) \mathfrak{d}(k)-\frac{%
\hat{v}\left( k\right) }{U}\mathfrak{s}\right\Vert _{2}\leq D_{\kappa
}\left( \epsilon \left\vert \hat{v}\left( k\right) \right\vert
^{-1}+U^{-2}\right) \ .
\end{equation*}
\end{proposition}

\noindent \textit{Proof.} By (\ref{v effec}) for $\gamma _{f}=\mathrm{e}%
^{\kappa }/2$ and (\ref{A legall0}), note that
\begin{equation*}
\tilde{A}_{1,1}(k)=A_{1,1}(k)-\frac{\mathrm{e}^{\kappa }}{2}\left( 1+4%
\mathrm{e}^{-2\kappa }\right) \left( \left\vert \hat{v}\left( k\right)
\right\vert -\frac{\mathrm{e}^{\kappa }}{4U+2}\left\vert \hat{v}\left(
k\right) \right\vert ^{2}\right) P_{\mathfrak{d}(k)}\ .\text{ }
\end{equation*}%
Therefore, similar to Theorem \ref{pairing mode}, it suffices to study the
operator
\begin{equation*}
UP_{0}-\frac{\mathrm{e}^{\kappa }}{2}\left( 1+4\mathrm{e}^{-2\kappa }\right)
\left( \left\vert \hat{v}\left( k\right) \right\vert -\frac{\mathrm{e}%
^{\kappa }}{4U+2}\left\vert \hat{v}\left( k\right) \right\vert ^{2}\right)
P_{\mathfrak{d}(k)}
\end{equation*}%
for large $U\geq 0$. Explicit computations shows in this case that the
negative eigenvalue of the last operator is%
\begin{eqnarray}
\mathrm{\tilde{F}}(k) &:=&\left( \frac{\left\vert \hat{v}\left( k\right)
\right\vert }{2}-\frac{\mathrm{e}^{\kappa }}{8U+4}\left\vert \hat{v}\left(
k\right) \right\vert ^{2}\right) \left( \frac{U}{\left\vert \hat{v}\left(
k\right) \right\vert -\frac{\mathrm{e}^{\kappa }}{4U+2}\left\vert \hat{v}\left( k\right) \right\vert ^{2}}-\left( \frac{\mathrm{e}^{\kappa }}{2}+2\mathrm{e}^{-\kappa }\right) \right.   \notag \\
&&\left. -\sqrt{\left( \frac{U}{\left\vert \hat{v}\left( k\right)
\right\vert -\frac{\mathrm{e}^{\kappa }}{4U+2}\left\vert \hat{v}\left(
k\right) \right\vert ^{2}}+2\mathrm{e}^{-\kappa }-\frac{\mathrm{e}^{\kappa }}{2}\right) ^{2}+4}\right)   \label{E-}
\end{eqnarray}%
with associated eigenvector $\tilde{\Phi}_{\uparrow \downarrow }(k)$ equal to%
\begin{equation}
\tilde{\Phi}_{\uparrow \downarrow }(k)=\left( \frac{\left\vert \hat{v}\left(
k\right) \right\vert }{-\mathrm{\tilde{F}}(k)}+\frac{\left\vert \hat{v}%
\left( k\right) \right\vert }{U}\right) \ \mathfrak{d}(k)+\frac{\left\vert
\hat{v}\left( k\right) \right\vert }{U}\mathfrak{s}\ .  \label{E-2}
\end{equation}%
Then, direct estimates from (\ref{E-})--(\ref{E-2}) imply Assertions
(i)--(ii). Note that, in order to use Kato's perturbation theory as in
Theorem \ref{pairing mode} we need the estimate $\mathrm{\tilde{E}}(k)=%
\mathcal{O}\left( 1\right) $, uniformly with respect to $U\geq 0$, which is
deduced from (\ref{estimate}) as in Theorem \ref{effective BCS copy(3)} for $%
\mathrm{E}(k)$.\hfill $\Box $

\section{Appendix\label{appendix}}

\subsection{Direct Integral Decomposition\label{sect direct decomposition}}

For more details, we refer to \cite[Section XIII.16]{RS4}.

Let $(\mathfrak{M},\mathfrak{m})$ be any $\sigma $--finite measure space and
$(\mathcal{H},\left\langle \cdot ,\cdot \right\rangle _{\mathcal{H}})$ any
separable Hilbert space. The constant fiber direct integral%
\begin{equation*}
\int_{\mathfrak{M}}^{\oplus }\mathcal{H}\text{ }\mathfrak{m}(\mathrm{d}x)\
\end{equation*}%
is denoted by $L^{2}(\mathfrak{M},\mathfrak{m};\mathcal{H})$ and corresponds
to the usual Hilbert space of $\mathcal{H}$--valued functions on $\mathfrak{M%
}$ with scalar product%
\begin{equation*}
\left\langle f,g\right\rangle :=\int_{\mathfrak{M}}\left\langle
f(x),g(x)\right\rangle _{\mathcal{H}}\mathrm{\ }\mathfrak{m}(\mathrm{d}x)%
\text{ }.
\end{equation*}

Recall that we denote the Banach space of bounded operators acting on $%
\mathcal{H}$ by $\mathcal{B}(\mathcal{H})$ with operator norm $\Vert \cdot
\Vert _{\mathrm{op}}$. A map $A(\cdot )$ from $\mathfrak{M}$ to $\mathcal{B}(%
\mathcal{H})$ is called \emph{measurable} if and only if the map $x\mapsto
\left\langle \psi _{1},A(x)\psi _{2}\right\rangle $ from $\mathfrak{M}$ to $%
\mathbb{R}$ is measurable for all $\psi _{1},\psi _{2}\in \mathcal{H}$.

Let $L^{\infty }(\mathfrak{M},\mathfrak{m};\mathcal{B}(\mathcal{H}))$ be the
space of equivalence classes of measurable functions $A:\mathfrak{M}%
\rightarrow \mathcal{B}(\mathcal{H})$ with%
\begin{equation*}
\left\Vert A\right\Vert _{\infty }:=\mathrm{ess}\text{ }\sup \left\Vert
A(\cdot )\right\Vert _{\mathrm{op}}<\infty \ .
\end{equation*}

A bounded operator $A$ on $L^{2}(\mathfrak{M},\mathfrak{m};\mathcal{H})$ is
\emph{decomposable} or \emph{decomposed} by the direct integral
decomposition if and only if there is $A(\cdot )\in L^{\infty }(\mathfrak{M},%
\mathfrak{m};\mathcal{B}\mathfrak{(}\mathcal{H)})$ such that%
\begin{equation*}
\left( A\Psi \right) \left( x\right) =A\left( x\right) \Psi \left( x\right)
\ ,\qquad \Psi \in L^{2}(\mathfrak{M},\mathfrak{m};\mathcal{H})\ .
\end{equation*}%
\ The operators $A\left( x\right) \in \mathcal{B}(\mathcal{H})$ are the
so--called fibers of $A$ and we write%
\begin{equation*}
A=\int_{\mathfrak{M}}^{\oplus }A(x)\text{ }\mathfrak{m}(\mathrm{d}x)\ .
\end{equation*}

The space of decomposable operators can be isometrically identified with the
space $L^{\infty }\left( \mathfrak{M},\mathfrak{m};\mathcal{B}(\mathcal{H}%
)\right) $. See, e.g., \cite[Theorem XIII.83]{RS4}. \cite[Theorem XIII.85]%
{RS4} also gives properties of self--adjoint operators on the space $%
L^{\infty }\left( \mathfrak{M},\mathfrak{m};\mathcal{B}(\mathcal{H})\right) $%
\ in terms of its fibers. Only \cite[Theorem XIII.85 (a), (c), (d)]{RS4} is
used in this paper and so, for the reader's convenience, we \ reproduce it
below:

\begin{proposition}[Self--adjoint decomposition]
\label{proposition Self--adjoint decomposable operators}\mbox{ }\newline
Let $A\in L^{\infty }\left( \mathfrak{M},\mathfrak{m};\mathcal{B}(\mathcal{H}%
)\right) $ with $A\left( x\right) $ being self--adjoint for any $x\in
\mathfrak{M}$. Then:\newline
\emph{(i)} $A$ is self--adjoint with spectrum $\sigma (A)$. \newline
\emph{(ii)} $\lambda \in \sigma (A)$ if and only if, for all $\varepsilon >0$%
,%
\begin{equation*}
\mathfrak{m}\left( \left\{ x\in \mathfrak{M}:\sigma (A(x))\cap (\lambda
-\varepsilon ,\lambda +\varepsilon )\neq \emptyset \right\} \right) >0\ .
\end{equation*}%
\emph{(iii)} For any bounded Borel function $f$ on $\mathbb{R}$,
\begin{equation*}
f\left( A\right) =\int_{\mathfrak{M}}^{\oplus }f\left( A(x)\right) \text{ }%
\mathfrak{m}(\mathrm{d}x)\ .
\end{equation*}
\end{proposition}

\subsection{The Birman--Schwinger Principle}

There are various versions of the Birman--Schwinger principle. The following
one is used in our proofs:

\begin{proposition}[Birman--Schwinger principle]
\label{lem-2.2}\mbox{ }\newline
Let $d\geq 1$ and $H_{0},V\in \mathcal{B}(\ell ^{2}(\mathbb{Z}^{d}))$ be
positive bounded operators. Assume further that $V$ is compact. For any $%
\lambda <0$, define the compact, self-adjoint, positive \emph{%
Birman--Schwinger operator} by
\begin{equation*}
B(\lambda )\ =\ B(\lambda ,H_{0},V)\ :=\ V^{1/2}\,(H_{0}-\lambda
)^{-1}\,V^{1/2}\ .
\end{equation*}%
Then $\lambda <0$ is an eigenvalue of $(H_{0}-V)$ of multiplicity $M$ if and
only if $1$ is an eigenvalue of $B(\lambda )$ of multiplicity $M$.
\end{proposition}

\noindent \textit{Proof}: We recall that, due to the compactness of $V$, the
Birman-Schwinger operator $B(\lambda )$ is compact and has only discrete
spectrum above $0$. Similarly, the spectrum of $(H_{0}-V)$ below $0$ is
discrete because $V$ is compact.

Suppose that $\lambda <0$ is an eigenvalue of $(H_{0}-V)$ of multiplicity $%
M\in \mathbb{N}$ and let $\{{\varphi }_{1},\ldots ,{\varphi }_{M}\}\subseteq
\ell ^{2}(\mathbb{Z}^{d})$ be an orthonormal basis (ONB) of the
corresponding eigenspace. Set
\begin{equation}
\psi _{1}:=V^{1/2}{\varphi }_{1},\ldots ,\psi _{M}:=V^{1/2}{\varphi }_{M}\in
\ell ^{2}(\mathbb{Z}^{d})\ .  \label{eq-2.4}
\end{equation}%
Then
\begin{equation}
{\varphi }_{m}=(H_{0}-\lambda )^{-1}V{\varphi }_{m}=(H_{0}-\lambda
)^{-1}V^{1/2}\psi _{m}\ ,  \label{eq-2.5}
\end{equation}%
and the boundedness of $(H_{0}-\lambda )^{-1}V^{1/2}$ implies that $\{\psi
_{1},\ldots ,\psi _{M}\}\subseteq \ell ^{2}(\mathbb{Z}^{d})$ is a linearly
independent family. Clearly, (\ref{eq-2.4}) and (\ref{eq-2.5}) also yield%
\begin{equation}
B(\lambda )\psi _{m}=V^{1/2}(H_{0}-\lambda )^{-1}V^{1/2}\psi _{m}=\psi _{m}\
,  \label{eq-2.6}
\end{equation}%
and hence the eigenspace of $B(\lambda )$ corresponding to the eigenvalue $1$
has at least dimension $M$.

Conversely, if $\{\psi _{1},\ldots ,\psi _{L}\}\subseteq \ell ^{2}(\mathbb{Z}%
^{d})$ is an ONB of the eigenspace of $B(\lambda )$ corresponding to the
eigenvalue $1$ of multiplicity $L\in \mathbb{N}$ then we set
\begin{equation}
{\varphi }_{1}:=(H_{0}-\lambda )^{-1}V^{1/2}\psi _{1},\ldots ,{\varphi }%
_{L}:=(H_{0}-\lambda )^{-1}V^{1/2}\psi _{L}\in \ell ^{2}(\mathbb{Z}^{d})\ .
\label{eq-2.7}
\end{equation}%
Then,
\begin{equation}
\psi _{k}=B(\lambda )\psi _{k}=V^{1/2}{\varphi }_{k}\ ,  \label{eq-2.8}
\end{equation}%
and the boundedness of $V^{1/2}$ implies that $\{{\varphi }_{1},\ldots ,{%
\varphi }_{L}\}\subseteq \ell ^{2}(\mathbb{Z}^{d})$ is a linearly
independent family. Clearly, (\ref{eq-2.7}) and (\ref{eq-2.8}) also yield
\begin{equation}
(H_{0}-V){\varphi }_{k}=\lambda {\varphi }_{k}\ ,  \label{eq-2.9}
\end{equation}%
and hence the eigenspace of $(H_{0}-V)$ corresponding to the eigenvalue $%
\lambda $ has at least dimension $L$.\hfill $\Box $

\bigskip

\noindent \textit{Acknowledgments:} This research is supported by the agency
FAPESP under Grant 2013/13215-5 as well as by the Basque Government through
the grant IT641-13 and the BERC 2014-2017 program and by the Spanish
Ministry of Economy and Competitiveness MINECO: BCAM Severo Ochoa
accreditation SEV-2013-0323.


\begin{thebibliography}{99}
\bibitem{Bonn} Bonn, D. A.: Are high-temperature superconductors exotic?
Nature Physics \textbf{2}, 159--168 (2006)

\bibitem{BruPedra1} Bru, J.B., de Siqueira Pedra, W.: Effect of a locally
repulsive interaction on s--wave superconductors. Rev. Math. Phys. \textbf{22%
}(3), 233--303 (2010)

\bibitem{BruPedraAniko} Bru, J.B., de Siqueira Pedra, W., D{\"{o}}mel, A.: A
microscopic two--band model for the electron-hole asymmetry in high-$T_{c}$
superconductors and reentering behavior. J. Math. Phys. \textbf{52},
073301-- (1-28) (2011)

\bibitem{Saxena} Saxena, A. K.: High-Temperature Superconductors.
Springer-Verlag, Berlin Heidelberg, (2010)

\bibitem{Rez} Reznik, D.: Giant Electron--Phonon Anomaly in Doped $%
La_{2}CuO_{4}$ and Others Cuprates. Advances in Condensed Matter Physics.
\textbf{2010}, 523549 (2010)

\bibitem{28} Pintschovius, L., Reichhardt, W.: Phonon Dispersions and Phonon
Density--of--States in Copper--Oxide Superconductors. In: Neutron Scattering
in Layered Copper--Oxide Superconductors, Furrer, A. editor. Physics and
Chemistry of Material with Low--Dimensional Structures, 165--223, Kluwer
Academic Publisers, Dordrecht (1998)

\bibitem{29} Ishihara, S., Egami, T., Tachiki, M.: Electron--Lattice
Interaction in Cuprates: Effect of Electron Correlation. Phys. Rev. B
\textbf{55}(5), 3163--3172 (1997)

\bibitem{61} Zhang, C. J., Oyanagi, H.: Local lattice instability and
superconductivity in $La_{1.85}Sr_{0.15}Cu_{1-x}M_{x}O_{4}$ ($M$=$Mn,Ni$ and
$Co$). Phys. Rev. B \textbf{79}(6), 064521 (2009)

\bibitem{62} Martelj, T., Kabanov, V. V., Mihailovic, D.: Charged particles
on a two--dimensional lattice subjected to anisotropic Jahn--Teller
interactions. Phys. Rev. Lett. \textbf{94}(14), 147003 (2005)

\bibitem{63} Miranda, J., Mertelj, T., Kabaniov, V. V., Mihailovic, D.:
Bipolaron Jahn--Teller pairing and charge transport in cuprates. Journal of
Superconductivity and Novel Magnetism \textbf{22}(3), 281--285 (2009)

\bibitem{64} Ranninger, J., Romano, A.: Local dynamical lattice
instabilities: prerequisites for resonant pairing superconductivity. Phys.
Rev. B \textbf{78}(5), 054527 (2008)

\bibitem{65} Keller, H., Bussmann--Holder, A., M{\"{u}}ller, K. A.:
Jahn--Teller physics and high--$T_{c}$ superconductivity. Materials Today
\textbf{11}(9), 38--46 (2008)

\bibitem{Ranninger} Ranninger, J.: The Polaron Scenario for High--$T_{c}$
Superconductors. In: Polarons and Bipolarons in High-Tc Superconductors and
Related Materials. Salje, E. K. H., Alexandrov. A. S., Liang, W. Y. editors.
Cambridge Univ. Press (1995)

\bibitem{1-0} Alexandrov, A. S., Ranninger, J.: Theory of Bipolarons and
Bipolaronic Bands. Phys. Rev. B \textbf{23}, 1796--1801 (1981)

\bibitem{1-1} Alexandrov, A. S., Ranninger, J.: Bipolaronic
Superconductivity. Phys. Rev. B \textbf{24}, 1164--1169 (1981)

\bibitem{58} Alexandrov, S. A.: Unconventional Pairs Glued by Conventional
Phonons in Cuprates. Journal of Superconductivity and Novel Magnetism
\textbf{22}(2), 103--107 (2009)

\bibitem{BZ1} Bru, J.-B., Zagrebnov, V.A.: Quantum Interpretation of
Thermodynamic Behaviour of the Bogoliubov Weakly Imperfect Bose-Gas, Phys.
Lett. A \textbf{247}, 37--41 (1998)

\bibitem{BruZagrebnov8} Zagrebnov, V.A., Bru, J.-B.: The Bogoliubov Model of
Weakly Imperfect Bose Gas. Phys. Rep. \textbf{350}, 291--434 (2001)

\bibitem{bru3} Bru, J.-B.: Beyond the dilute Bose gas, Physica A \textbf{359}%
, 306--344 (2006)

\bibitem{BruPedra2} Bru, J.B., de Siqueira Pedra, W.: Non--cooperative
Equilibria of Fermi Systems With Long Range Interactions. Memoirs of the
AMS. \textbf{224}(1052) (2013)

\bibitem{Pedra-Salmhofer} de Siqueira Pedra, W., Salmhofer M.: Determinant
Bounds and the Matsubara UV Problem of Many-Fermion Systems, Commun. Math.
Phys. \textbf{282}, 797--818 (2008)

\bibitem{meissner} Bru, J.-B., de Siqueira Pedra, W.: Microscopic
Foundations of the Mei{\ss }ner Effect -- Thermodynamic Aspects. Rev. Math.
Phys. \textbf{25}(7), 1350011-- (1-66) (2013)

\bibitem{RS4} Reed, M., Simon, B.: {Methods of Modern Mathematical Physics,
Vol. IV}. Academic Press, New York-London, (1972)
\end{thebibliography}
\end{document}